\documentclass[twocolumn]{aastex63}

\submitjournal{\apj}

\shorttitle{Carbon and Oxygen AGB stars in BAaDE}
\shortauthors{Lewis et al.}

\begin{document}

\title{Carbon- and Oxygen-rich Asymptotic Giant Branch (AGB) stars in the Bulge Asymmetries and Dynamical Evolution (BAaDE) survey}

\correspondingauthor{Megan O. Lewis}
\email{melewis@unm.edu}

\author{Megan O. Lewis}
\affiliation{University of New Mexico, Department of Physics and Astronomy \\
Albuquerque, NM 87131, USA}
\affiliation{National Radio Astronomy Observatory \\
Array Operations Center\\
Socorro, NM 87801, USA}

\author{Ylva M. Pihlstr\"om}
\altaffiliation{Ylva Pihlstr\"om is also an Adjunct Astronomer at the\\ National Radio Astronomy Observatory.}
\affiliation{University of New Mexico, Department of Physics and Astronomy \\
Albuquerque, NM 87131, USA}

\author{Lor\'ant O. Sjouwerman}
\affiliation{National Radio Astronomy Observatory \\
Array Operations Center\\
Socorro, NM 87801, USA}

\author{Michael C. Stroh}
\affiliation{Northwestern University, Department of Physics and Astronomy \\
Evanston, IL 60208, USA}

\author{Mark R. Morris}
\affiliation{University of California,
 Department of Physics and Astronomy\\
 Los Angeles, CA 90095, USA}

\author{the BAaDE Collaboration}


\begin{abstract}


Detections of SiO masers from the Bulge Asymmetries and Dynamical Evolution (BAaDE) survey more tightly define the region where Oxygen-rich (O) Asymptotic Giant Branch (AGB) stars reside in multiple infrared (IR) color-color diagrams. Using MSX and 2MASS data along with radio spectra from the BAaDE survey we find that three main populations were observed in the BAaDE survey: O-rich AGB stars of which about $73\%$ host SiO masers, Carbon-rich (C) AGB stars which do not host these masers, and a small contaminating set of possible Young Stellar Objects (YSOs). The distinction between YSOs and AGB stars can be drawn using only MSX data, specifically the $[D]-[E]$ color, while the shorter wavelengths provided by 2MASS are necessary to divide potential C- and O-rich AGB stars. Divisions similar to these have been seen in multiple earlier IR-studies, but BAaDE currently provides a sample of $\sim$ 15000 sources which far exceeds previous studies in sample size, and therefore provides much more distinct divisions. With these IR distinctions in place, we discuss the sources which are exceptions in either their molecular detections or IR colors, as well as the distribution of the three populations in Galactic coordinates.

\end{abstract}



\section{Introduction}\label{intro}
Stars between about 0.8 and 8 $M_\sun$ evolve through the Asymptotic Giant Branch (AGB) phase during which they undergo substantial mass loss. In their resulting dense circumstellar envelopes (CSEs) molecules form, with abundances reflecting whether the source is oxygen-rich (O) or carbon-rich (C). The majority of AGB stars are O-rich with a C/O ratio smaller than 1, and under these conditions SiO molecules easily form. However, the more massive of these AGB stars can experience a third dredge-up phase, and as long as hot-bottom burning does not occur C can be brought up to the stellar envelope via large convection cells (e.g.\,\citealt{dicris}). If a star has undergone this phase, the stellar atmosphere can contain more C than O, allowing the creation of a large range of C-bearing molecules \citep{tsuji,herwig}. Under this paradigm, O-rich AGB stars with certain CSE conditions will host SiO emission (commonly masers) while C-rich AGB stars will not. This is not always the case, as the well-known C-AGB IRC+10216 is known to display weaker (thermal) emission from SiO as well as other oxides \citep{twostars}. However in general, O-rich AGB stars show brighter emission from O-bearing molecules and, comparably, C-rich AGB stars show brighter emission from C-bearing molecules \citep{buj}.

\par Studies of AGB chemistry have focused on near, bright (and potentially atypical) sources such as the C-rich IRC+10216 and the O-rich IK Tauri \citep{twostars, thatthingylvagaveme}. Molecular detections in these sources imply that, in particular, C-stars have a rich chemistry. 
\par Samples of stars have also been studied to determine distributions of C/O-rich objects in the Galaxy, and to statistically identify the differences between O- and C-rich CSEs. These larger samples are typically collected from preexisting catalogues from the literature, gathering their O- and C- stars from separate sources \citep{loup, ortiz, lumsden, suh}. Other infrared (IR) studies such as Ishihara et al.\,\citeyearpar{ishi} rely on the literature to identify C- and O-rich AGB stars to subsequently establish IR-based color-cuts separating C-rich sources from O-rich ones. Both single-star and larger-sample data suggest that the differences between O- and C-rich dust properties cause differences in mass-loss histories, IR-colors, and CSE properties for O- and C-AGBs.

\par Being enshrouded in circumstellar dust, AGB stars are difficult to study at optical wavelengths and are more readily observable both at IR wavelengths and through their radio-wavelength molecular emission. As such, they have been the targets of numerous IR and maser surveys. One such survey is the BAaDE (Bulge Asymmetries and Dynamical Evolution) project in which approximately 19000 IR-selected targets have been observed for SiO maser emission with the NSF's Karl G. Jansky Very Large Array (VLA), and another $\sim$9000 targets are in the process of being observed with Atacama Large Millimeter/sub-millimeter Array (ALMA; Sjouwerman et al.\,2020, in preparation). This survey, and others like it, can be used to derive the line-of-sight velocities of the sources and thereby infer Galactic dynamics \citep{trapp}. BAaDE also provides a large sample of AGB stars with which to study maser brightnesses, maser pumping, and the distinctions between C- and O-rich AGB stars. 
For example, detections of SiO masers are known to trace the O-rich population \citep{habing}. While non-detections may still be O-rich and not detected due to variability or sensitivity \citep{ylva, michael}, a fraction of the non-detections may instead be associated with C-rich stars. By including IR data, the BAaDE data can be used to distinguish O- and C-rich AGB stars. Applying such a distinction for the close to 28000 stars in the BAaDE sample allows for exploration of the ratio of C-AGBs to O-AGBs (which is an indirect measure of metallicity; \citealt{metal}) as a function of Galactic location. It can further be used for improved estimates of the mass-return to the Galaxy by C-AGB stars via mass-loss estimates based on IR color (e.g., \citealt{claussen}, \citealt{vdVR}) and spectral energy distribution modeling which is in progress for the BAaDE sample \citep{eddie}. Identifying large numbers of C-AGBs is also of interest as these sources provide a large amount of processed material to the ISM \citep{epchtein}, but there seems to be a dearth of such sources in the Bulge \citep{noguchi} and the origin of the few Bulge C-AGBs is unknown \citep{matsunaga}.

 \par Making C/O distinctions based on IR-color aided by SiO maser detections/nondetections, as opposed to previous identifications from the literature, opens up possibilities for huge, uniform studies of the C- versus O-rich types. In this work, the detection rate of SiO masers as a function of IR color is used to determine the IR color ranges where O-rich AGB stars are found, and by extension to identify C- and O-rich AGB stars throughout the Galaxy based on their IR colors. Similar IR-color ranges have been found in the past (\citealt{ortiz}, \citealt{lumsden}, \citealt{suh}, \citealt{ishi}), but the size of the sample discussed here (currently 14742 sources) makes for clearer IR cuts between populations. 

\par Section \ref{obssect} discusses both the IR selection and spectral line observations made in the BAaDE survey. In Sect.\,\ref{resultsect} we lay out several IR color cuts that can be applied to separate O-AGBs, C-AGBs, and non-AGBs, and in Sect.\,\ref{discussionsect} we apply these cuts to the entire BAaDE sample. Section\,\ref{conclusion} concludes.

\section{Observations/BAaDE}\label{obssect}
The BAaDE project is the largest ever SiO maser survey covering approximately 28000 sources along the entire Galactic plane between about $-$5$^{\circ}$ and 5$^{\circ}$ Galactic latitude. The targets were chosen from the MSX point source catalog, using color selections indicative of O-rich AGBs, and therefore the targets have a high expected SiO maser detection rate (\citealt{scc}; hereafter SCC09). About 70\% (19000) of the sources have been observed with the VLA, with a spectral setup covering the v=0, 1, 2, and 3 $J=1-0$ transitions near 43 GHz (as well as transitions from the $^{29}$SiO and $^{30}$SiO isotopologues). The other 30\% of sources are not observable from the VLA and so are being observed by ALMA covering the v=0, 1, and 2 $J=2-1$ transitions at 86 GHz (as well as a transition from $^{29}$SiO). Currently, about 1400 ALMA sources have been observed. The effects of observing with two distinct instruments and at two different frequencies are discussed in Stroh et al.\,\citeyearpar{michael}. Both Sjouwerman et al.\,\citeyearpar{sjouwer} and Stroh et al.\,\citeyearpar{michael} find that stars with 43 GHz SiO masers generally also host 86 GHz SiO masers and vice versa. Additionally, Stroh et al.\,\citeyearpar{michael} find that 43 GHz masers are, on average, slightly brighter. The ALMA 86 GHz portion of the sample comprising the far side of the bar is compensated by a higher ALMA sensitivity per source, so that distance dominated sensitivity effects are minimal. In addition, in the study presented in this paper only $\sim$10\% of the data come from the ALMA 86GHz observations and thus comprise a minority of the discussed sample. The ALMA data is nevertheless important to include as it provides key information via the occasional detection of a CS line, which is discussed in Sect.\,\ref{ysodr} and \ref{ysoim}.
The VLA data were collected between 2013-2017 with a spectral resolution of 1.7 km\,s$^{-1}$ and a typical root-mean-squared noise level of about 15-20 mJy/beam. A source is determined to be an SiO-detection if there is a 4.7-sigma (or higher) detection within 400 km\,s$^{-1}$ of the LSR-corrected rest frequency of an SiO transition. The partial sample of ALMA data used for this study was collected in 2015, 2016, and 2018 at a spectral resolution of 0.8 km\,s$^{-1}$ and to approximately the same noise level as the VLA data. Within the ALMA data single-line detections at the 5-sigma level are recorded along with multiple line detections at the 4.5-sigma level. For a more complete description of the BAaDE survey see \cite{baade}.
\par In this paper we utilize a collection of  14742 (13314 VLA and 1428 ALMA) sources that represent the sources for which the data have been analyzed to date. Spectra for each source can be found on the survey website  \footnote{http://www.phys.unm.edu/$\sim$baade/specs/table.shtml}. The initial observed SiO maser detection rate is $\sim57\%$ for the VLA sources and $\sim72\%$ for the ALMA sources. The overall (VLA and ALMA) detection rate rises to 73\% when contaminants like Young Stellar Objects (YSOs) are removed (see also Sect.\,\ref{detrate}). It should be noted that the detection rates vary depending on position on the sky, which is partly due to the efficiency of the calibration strategy (Sjouwerman et al. 2020, in preparation), but also due to the uneven depths the Midcourse Space eXperiment (MSX) survey observed at different longitudes (\citealt{msxpsc}; see also Sect.\,\ref{detrate}). The only molecule identified in any of the VLA spectra so far is SiO, while we have identified CS, H$^{13}$CN, and H$^{13}$CO$^{+}$ in addition to SiO in the ALMA data. The lack of additional molecules in the VLA data can be attributed to the different spectral coverages of the ALMA and the VLA observations. As a consequence, we can use SiO maser emission as an indicator of O-rich AGB stars, but we mostly rely on low detection rates of SiO rather than on a detection of a carbon-bearing molecule to identify other possible populations within the sample. CS detections can be used, to some extent, to identify other populations in the ALMA data; they are associated with a small population of YSOs as well as with candidate C-AGBs (see Sect.\,\ref{ysosect}). A single CS line is therefore not a reliable tracer of any specific population in the BAaDE survey and does not aid in identifications across the entire survey unless other data, like an IR color, are added.
\par This work relies primarily on the currently analyzed sources from the BAaDE sample (referred to as 'this sample' or 'our sample' in the text); however, once correlations between SiO-detection rates and IR-colors are established, many of our results can be extended to all BAaDE sources (referred to as 'the BAaDE sample' or 'the entire BAaDE sample') and/or all MSX point sources. 

\subsection{IR data}\label{irselectsect}

The sequence of regions in the IRAS color-color diagram presented by van der Veen and Habing (1988, their Fig.\,2) is commonly thought of as a sequence of increasing CSE optical-depth, and the regions laid out in that work are often used as the basis of IR color source selection (e.\,g., \citealt{tlh}). The sources in the BAaDE survey were selected to give a high chance of SiO maser activity, corresponding to section IIIa of the van der Veen and Habing IRAS color-color diagram, which is the region inhabited by variable AGB stars with dusty circumstellar shells and the 9.7 $\mu$m silicate feature in emission \citep{vdvhabing}. Unfortunately, the IRAS survey is highly confused in the crowded area of the Galactic plane, so the actual selection criteria used for BAaDE were determined by mapping sections of the IRAS color-color diagram to MSX colors (SCC09). In the mapping, each IRAS two-color region (designated by uppercase roman numerals) is assigned an MSX counterpart (designated by lowercase roman numerals). MSX has higher angular resolution (18\arcsec) than IRAS ($\sim$1\arcmin), allowing for a color-based target selection in the Galactic plane. BAaDE sources were chosen from the MSX Point Source Catalogue version 2.3 \citep{msxpsc} based on their MSX $[A]-[D]$, $[A]-[E]$, or $[C]-[E]$ color\footnote{MSX $[A]$, $[C]$, $[D]$, and $[E]$ bands correspond to center wavelengths of 8.28, 12.13, 14.65, and 21.34 $\mu$m, respectively} to belong to SCC09 region \textit{iiia} using the following criteria for uncorrected MSX magnitudes: $-0.6 < [A]-[D] < 0.4$, $-0.7 < [A]-[E] < 0.4$, or $-0.75 < [C]-[E] < 0.1$. A vast majority of sources (96.7\%) was selected by their $[A]-[D]$ value. However, as shown by SCC09, the relatively narrow range of wavelengths covered by MSX (8.28$-$21.34 $\mu$m) prevented a direct mapping from the IRAS color-color diagram. C-AGBs are preferentially identified using the longer 60$\mu$m IRAS wavelengths, and thus cannot easily be separated from O-rich AGB stars in the MSX data. The BAaDE sample has therefore not actively deselected C-AGBs, even though we do not expect SiO detections in these sources (SCC09).

\par All BAaDE sources have been cross-matched with the 2MASS, AKARI, WISE, and MIPSGAL IR surveys, with the most matches present in 2MASS (27050 2MASS matches, 24821 AKARI matches, 8925 WISE matches, and 12875 MIPSGAL mataches out of the 28062 total BAaDE sources). The cross-matching is discussed in Hilburn \citeyearpar{eddie} and follows the general scheme of searching within a specific radius which is dependent on the positional accuracies of the surveys, and assigning the reddest source as a match in the case of multiple source matches. The cross-matching for the full sample of 28026 stars is not completely finalized, because of, for example, insufficient reddening corrections, but at this time we estimate that less than 3\% of the cross-matches between MSX and 2MASS may be incorrect counterparts. For most sources in the BAaDE sample there thus exists a set of catalogued IR magnitudes in the wavelength range 1.2-24$\mu$m, as well as known maser line fluxes and ratios. The work presented here relies on MSX data as well as 2MASS [K$_s$] magnitudes (centered at 2.2$\mu$m) and AKARI [18] magnitudes (centered at 18$\mu$m) that were obtained through this cross-matching system. In order to compare these magnitudes across studies, zero-magnitude flux densities of 55.810 Jy, 26.423 Jy, 18.267 Jy, 8.666 Jy, 665 Jy, and 12.001 Jy are applied to MSX $[A]$, $[C]$, $[D]$, $[E]$, 2MASS $[K_s]$, and AKARI [18] magnitudes respectively (SCC09; \citealt{2mass}; \citealt{akari}). No other corrections (i.e., reddening corrections) were applied to any of the IR magnitudes or colors used in this work.

\section{Results}\label{resultsect}
The goal of the work presented in this paper is to characterize the BAaDE sample in order to ensure more nuanced interpretations of the data. The MSX IR selection process is designed to favor O-rich stars, but it does not directly allow a deselection of C-rich stars, which were expected to be a few percent of the sample. Further, the IR color selection does not prevent a small fraction of YSOs from entering the sample. If kinematic populations are to be distinguished, YSOs must be filtered out from the more evolved sources. SiO maser detections imply O-rich AGB stars which should be the bulk of our sample. By including line detections from C-bearing molecules and IR data, we show that YSOs and candidate C-rich AGB stars can be identified as well.

\subsection{Removing non-AGB sources: YSO population}\label{ysosect}
The initial IR-selection for the BAaDE survey contained color ranges for luminous sources embedded in dusty material where AGB stars, but also YSOs, can be found. Here we show that the YSO population in our sample is easily identified by its MSX $[D]-[E]$ color, with `DE-red' sources ($[D]-[E]>$1.35) comprising the YSOs. This population differentiates itself from the 'DE-blue' sample in several ways: 1) YSOs occupy a distinct space in various IR color-color diagrams (Sect.\,\ref{ysocc}), 2) they are spatially clumped in and around molecular clouds in the plane (Sect.\,\ref{ysosd}), 3) their spectra show a lack of SiO detections and a high rate of CS detections (Sect.\,\ref{ysodr}), and 4) when imaged, the CS emission is often extended rather than point-like (Sect.\,\ref{ysoim}; \citealt{michael2}). It should be noted that although the positions of the DE-red sources in color-color diagrams and in the Galactic plane lead us to identify the population as dominated by YSOs, some sources may be post-AGB sources, planetary nebulae, red super-giants\footnote{Red super-giants are especially likely to be mistaken for YSOs, because not only do the  two classes share similar IR colors, they are also both dynamically young---meaning that the two are likely to share the characteristics of being positioned in the plane and near molecular clouds. As red super-giants form from rare massive stars and are short lived, we do not expect many of them in our sample and therefore the DE-red population is likely dominated by YSOs.}, or HII regions as there is overlap between these types of sources in IR color-color diagrams and not all of our DE-red sources are associated with clouds. The term 'YSO population' as used in the rest of this paper refers to the DE-red population even though we acknowledge that there may be other types of objects among the DE-red sources.

\begin{figure*}[tb]
    \centering
    \includegraphics[width=5in]{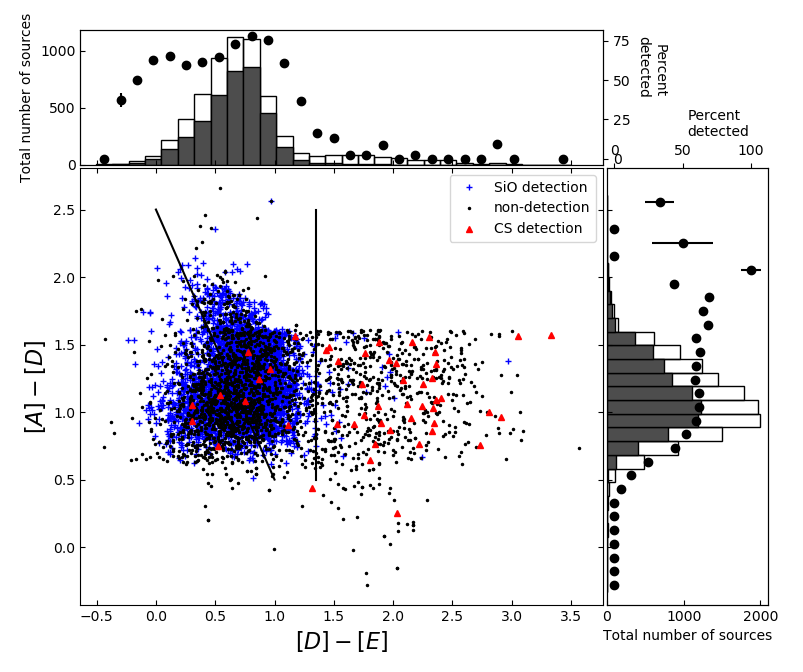}
    \caption{Scatter plot: Zero-magnitude-corrected color-color diagram of the sources in our sample with reliable $[E]$-magnitudes. DE-red and DE-blue sources are divided by the vertical line. DE-red sources are not AGB stars, as discussed in Sect.\,\ref{ysosect}, and show a low SiO maser detection rate. The OC-dividing line, discussed in Sect.\,\ref{ocindexsect}, which divides O- and C-rich AGB stars effectively in the Ortiz et al. \citeyearpar{ortiz} sample and defines the OC-index, is shown as the diagonal line to the left. This line does not effectively divide O- and C-rich objects in our sample, and the SiO detection rate is relatively high and constant across all DE-blue sources in this diagram (see also Fig.\,\ref{ocindex_old}). The sharp cut-offs in [A]$-$[D] values are artifacts of the selection criteria of the BAaDE survey. Edge plots: Histogram of the number of sources per color bin (white bars, corresponding to left and bottom axes), histogram of number of SiO detections per color bin (grey bars, corresponding to left and bottom axes), and the corresponding SiO detection rates as percentages (black dots, corresponding to right and top axes). Error bars on the detection rates are often smaller than the symbol size.}
    \label{fig:dead}
\end{figure*}

\subsubsection{IR color-color diagrams}\label{ysocc}
Lumsden et al.\,\citeyearpar{lumsden} use MSX [A]$-$[D] versus [D]$-$[E] diagrams to separate a sequence of AGB stars from YSOs. In our sample, 6656 sources (45\%) have reliable [E]-magnitudes and BAaDE spectra and so can be plotted in a diagram (Fig.\,\ref{fig:dead}) corresponding to that of Lumsden et al. (\citeyear{lumsden}, their Fig. 10). Although the Lumsden et al.\,\citeyearpar{lumsden} sample and our BAaDE sample were selected very differently, there are some common aspects to their distributions in this color-color diagram. Two populations can be seen in our data, separated at $[D]-[E]=1.35$. This value was determined by fitting a pair of Gaussians to the [D]$-$[E] values of all of the BAaDE sources with [E]-magnitudes, and determining where the Gaussian curves intersect\footnote{The mean and standard deviation of the fit to the DE-blue distribution are 0.67 and 0.27, and for the DE-red distribution, 1.92 and 0.53}. Comparison of our sample with that of Lumsden et al (2002; first panel of their figures 3, 4, and 5) indicates that `DE-blue' sources with $[D]-[E]<1.35$ are primarily O-rich AGB stars, while DE-red sources are primarily YSOs. 
\begin{figure}
    \centering
    \includegraphics[width=.45\textwidth]{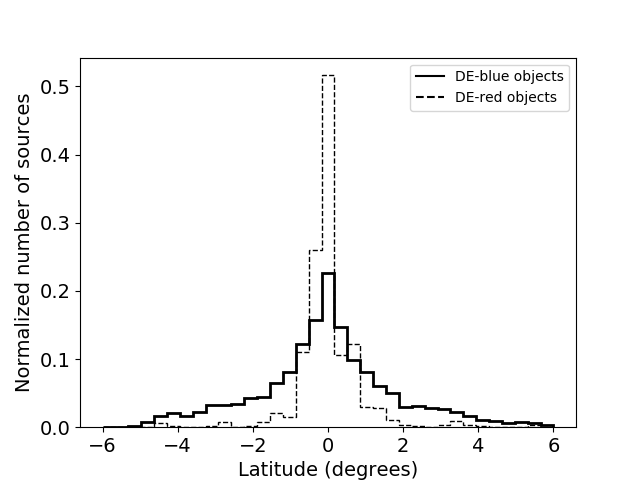}
    \caption{Distributions of BAaDE sources in Galactic latitude normalized to the number of sources. The solid line shows DE-blue objects which we identify as AGB sources; the dashed line shows DE-red objects which are likely YSOs. The narrow range of latitudes in the distribution of the DE-red objects supports their identification as primarily YSOs.}
    \label{lathist}
\end{figure}
\subsubsection{Spatial distribution}\label{ysosd}
\par
\begin{figure}[tb]
    \centering
    \includegraphics[width=.47\textwidth]{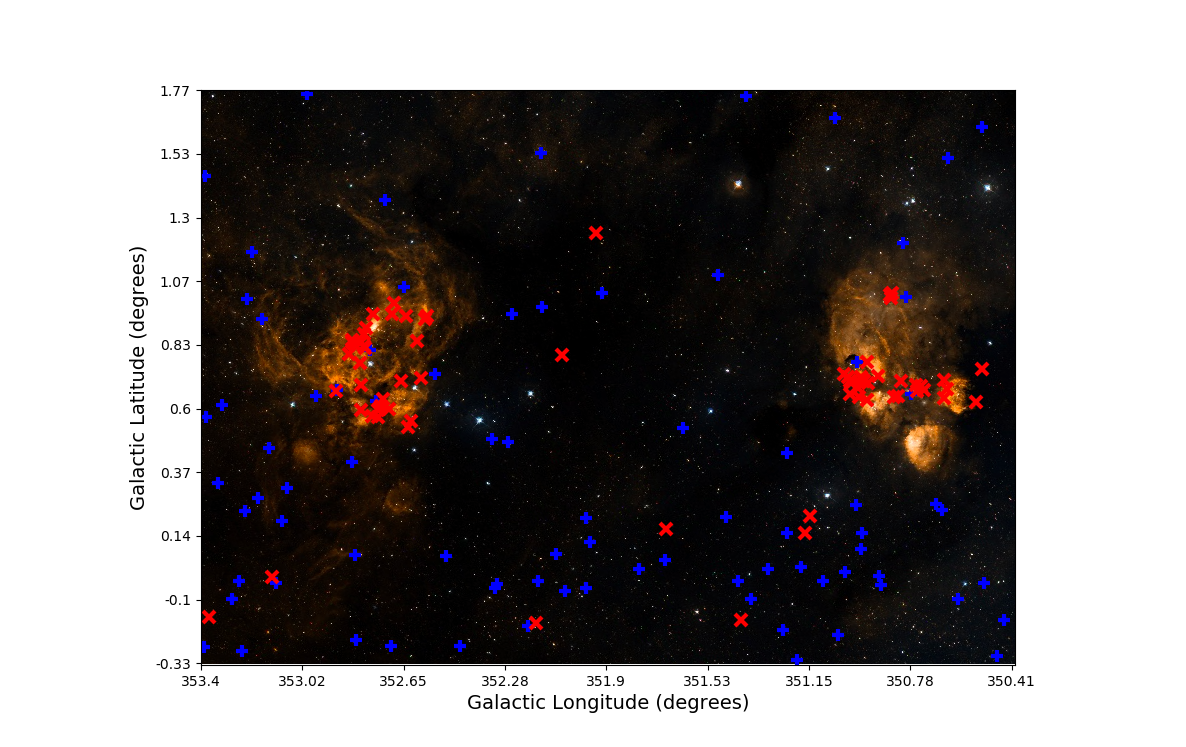}
    \caption{BAaDE sources within a 2$^{\circ}$ by 3$^{\circ}$ field of view centered on $l=352.2^{\circ}$ $b=0.7^{\circ}$. DE-red (red x) and DE-blue (blue +) sources are plotted on the Digital Sky Survey (DSS) image. The spatial distribution of the DE-red sources is often clumped around molecular clouds as shown in the figure, consistent with DE-red sources probably belonging to the YSO class.}
    \label{ysoclump}
\end{figure}

 Compared to the AGB sample, the DE-red sources are distributed in a much narrower latitude range close to the Galactic plane (Fig.\ref{lathist}). DE-red sources are also clumped in and around molecular clouds. An illustrative example of this is shown in Fig.\,\ref{ysoclump}, where, by placing DE-red and DE-blue sources on their corresponding Digital Sky Survey backdrop, it can be seen that the DE-red sources show a clear tendency to clump towards the molecular clouds while the DE-blue are more uniformly distributed. Both these patterns in the distribution of DE-red objects suggest that these objects were likely formed in molecular clouds in the plane and have not had time to disperse. Therefore, both the narrow latitude range and the spatial clustering coinciding with molecular clouds is in agreement with the classification of DE-red sources as primarily YSOs.

\subsubsection{Detection rate}\label{ysodr}
In addition to the DE division, Fig.\,\ref{fig:dead} also shows the significant differences in the number of molecular detections found in the two groups. The dots in the top panel of Fig.\,\ref{fig:dead} trace the SiO detection rate across color which drops steeply at the DE division. The SiO detection rate drops from 70\% on the left-hand side of the vertical line (low [D]$-$[E] values) to 5\% on the right-hand side, and CS detections are preferentially found on the right side of the line, consistent with the DE-blue population being O-rich AGB stars.

\subsubsection{CS emission distribution}\label{ysoim}
Molecular emission originating in AGB CSEs is expected to be centrally distributed around the star, and in particular SiO---which is found at a few stellar radii---can be expected to be unresolved in the BAaDE survey. To identify the nature of the CS detected in the BAaDE survey, Stroh et al. (2019) mapped the CS emission. The emission for the majority of the CS-detected DE-red sources was found to be extended, while the DE-blue sources displayed a much more compact distribution. While there is some overlap between the DE-red and DE-blue samples, the trend is strongly consistent with the DE color population separation.

\subsection{C-rich versus O-rich AGB populations}
\begin{figure*}[htb]
    \centering
    \includegraphics[width=5in]{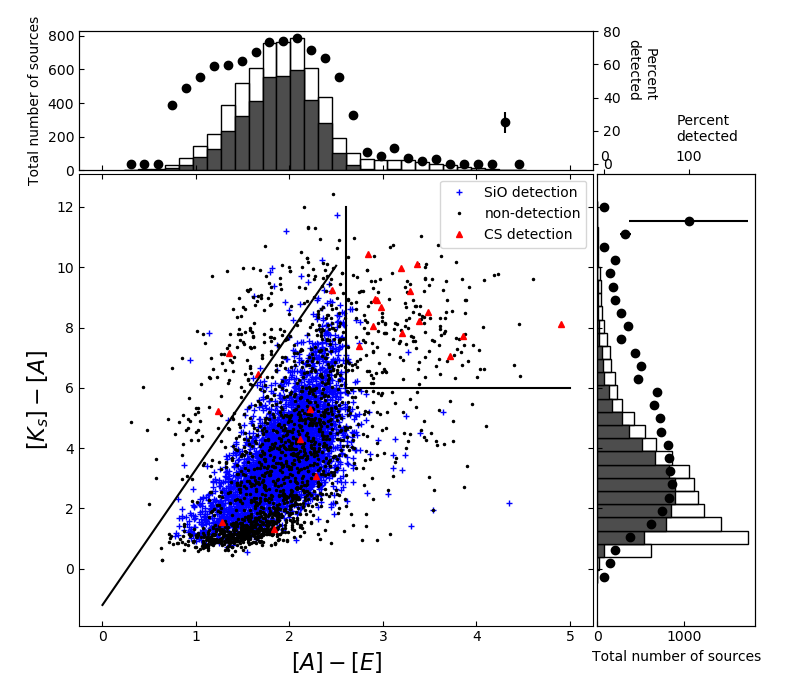}
    \caption{Scatter plot: Zero-magnitude corrected color-color diagram of the sources in our sample with reliable $[K_s]$ and $[E]$-magnitudes showing two areas of low SiO detection rates. One corresponds to the DE-red/non-AGB population discussed in Sect.\,\ref{ysosect} and the other to candidate C-AGBs. The box in the upper right roughly sections off the DE-red population for clarity; the population is not defined by the cuts shown here, but rather by $[D]-[E]$ color. The diagonal line defines our distinction between O-rich and C-rich AGB stars (see Eq. 1 and Fig.\,\ref{ocindex_new}). SiO detections are common in the O-AGB region, and CS detections are most common among the non-AGBs. SiO detections outside of the O-AGB region and CS detections within said region are not in keeping with our scheme and are discussed as sources of interest in Sect.\,\ref{exceptionsect}. Edge plots: Histogram of the number of sources per color bin (white bars, corresponding to left axes), histogram of the number of SiO detections per color bin (grey bars, corresponding to left axes), and the corresponding SiO detection rate (black dots, corresponding to right axes). Error bars on the detection rates are often smaller than the symbol size.}
    \label{fig:kaae}
\end{figure*}
\begin{figure}
    \centering
    \includegraphics[width=.47\textwidth]{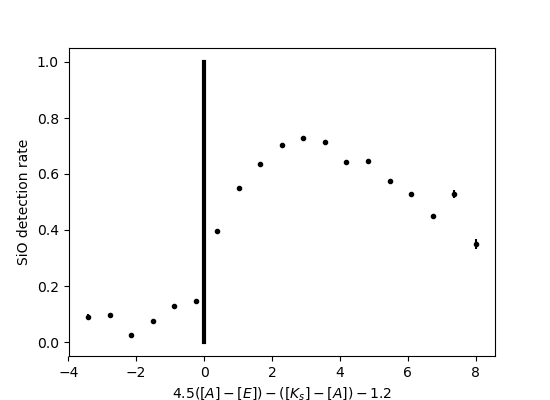}
    \caption{BAaDE SiO maser detection rate per 0.6 magnitude bin as a function of the distance from the line described in Equation 1 (shown in Fig.\,\ref{fig:kaae}). DE-red sources and bins with fewer than 10 sources have been removed. Error bars are often smaller than the symbol size. We propose this index is a good predictor of whether a source is O- or C-rich in the BAaDE data, where a negative index predicts a more likely C-rich source and positive predicts more likely O-rich; the solid line shows the division between the two. Our sample shows a steep decrease in detection rate for negative indices, which aids in identifying the negative sources as candidate C-AGBs.}
    \label{ocindex_new}
\end{figure}

With DE-red sources (mostly YSOs) removed from the sample we can discuss the AGB population in the survey with very little contamination from non-AGB objects. SiO detections in conjunction with IR-colors can also be used to separate C-rich from O-rich AGB stars. SCC09 found that one cannot divide O- and C-rich stars based solely on MSX colors because of the narrow total wavelength range covered by MSX. Our sample shows similar results with no clear division on MSX color-color diagrams (see Sect.\,\ref{ocindexsect}). Therefore, following Lumsden et al.\,\citeyearpar{lumsden}, we utilize both 2MASS and MSX colors to separate O- and C-rich AGB stars. We find, DE-red sources excluded, that there is a clear division between a population with a high SiO detection rate ($\sim$73\%) and one with a low rate ($\sim$12\%) in the $[K_s]-[A]$ versus $[A]-[E]$ color-color diagram (Fig.\,\ref{fig:kaae}). These populations can be interpreted to represent O-rich and C-rich AGB stars respectively. This division scheme closely matches the populations found in the same color-color space in Lumsden et al (2002; first panel of their figures 9, 10, and 11), with candidate C-rich AGB stars occupying a space that is redder in $[K_s]-[A]$ color and bluer in $[A]-[E]$ color than the O-rich AGB. The YSO population discussed in Sect.\,\ref{ysosect}, has not been removed from Fig.\,\ref{fig:kaae} to show that the YSOs occupy the reddest (in both colors) section of the diagram and are easy to visually separate from both C and O-AGBs, which also agrees with Lumsden et al.\,\citeyearpar{lumsden}. Empirically we define:
\begin{equation}
    [K_s]-[A]=4.5([A]-[E])-1.2 
\end{equation}
as the division between the populations. Based on this division the C-rich region is characterized more by a low detection rate of SiO than a high rate of any C-bearing molecule (see Fig.\,\ref{ocindex_new}). This is due mostly to the fact that VLA sources make up 90\% of the sources examined in this work, and that there are no certain detections of any molecule other than SiO in any VLA spectrum. We also lack IR spectroscopy, including the Si or C containing dust bands often used to identify C- versus O- rich stars. We do not assume that an SiO non-detection in the survey is solely indicative of a non-O-AGB source, as stellar variability, the sensitivity of our observations, and the fact that some O-AGB stars may not have CSE conditions conducive to maser activity all play a role.

\subsubsection{Candidate C-rich AGB stars}
\label{crichsect}
As seen by comparison of Fig.\,\ref{fig:kaae} with similar figures in Lumsden et al.\,\citeyearpar{lumsden}, the carbon-stars from Alksnis et al.\,\citeyearpar{alksnis} fall mostly into our C region. The color-color regions defined here for C- and O-rich AGB stars can also be checked using BAaDE sources with known identifications. For example, in the C-rich AGB region of Fig.\,\ref{fig:kaae} there are two previously identified C-stars with CS emission, \textit{ad3a-27052} and \textit{ad3a-25994}. These sources show several common traits: their IR-colors lie in the C-AGB region of Fig.\,\ref{fig:kaae}, they show only thermal SiO, CS, and H$^{13}$CN detections, the molecular lines are wide compared to most detections in BAaDE ($>$40 channels, 60 km\,s$^{-1}$), and the molecular emission is compact (images in Stroh et al.\,2019). Although not previously identified as a C-AGB, \textit{ad3a-26120} also shares these characteristics; therefore, we suggest \textit{ad3a-26120} is a strong candidate C-AGB, making all three sources with CS detections in the C-AGB region C-stars. Many non-detections in the C-AGB region may also belong to C-AGB sources whose molecular emission is below the sensitivity of the BAaDE survey, which was intended to detect maser emission. Candidate C-AGB sources observed by the VLA are especially likely to be non-detections as the spectral set-up for those observations was judiciously chosen to optimize coverage of multiple SiO lines as opposed to strong or common C-bearing molecular lines.   
\subsection{Sources which are exceptions}\label{exceptionsect}
Most CS detected objects lie either with the C-AGB or YSO populations, which is in keeping with the scheme that O-rich objects do not have many C atoms outside of CO molecules with which to form C-rich molecules like CS. Similarly SiO detected sources lie almost exclusively in the O-rich AGB section. The exceptions (CS detections in the O-AGB region and SiO detections outside of that region) are listed in Tables 1-3 with brief suggested explanations for each. It is not unexpected that our IR cuts sometimes show sources bleeding over into the wrong color-region because a) our IR-colors have not been reddening corrected, and b) the 2MASS and MSX data for these variable objects were not taken simultaneously leading to small uncertainties in any calculated color value. Any photometric errors in the 2MASS or MSX catalogues would also add to magnitude and color uncertainties. More detailed explanations follow.

\begin{deluxetable*}{ccccccc}[htb]
\tablehead{
\colhead{BAaDE name} & \colhead{Other name} & \colhead{RA and}  & \colhead{BAaDE} & \colhead{Explanation}
\\ \colhead{} & \colhead{} & \colhead{Dec}  & \colhead{morphology} & \colhead{}
}

\startdata
ad3a-22405 & IRAS 15585$-$5302 & 16:02:21.81 & extended CS, & S-star candidate\\
 &   & $-$53:10:52.9 & point SiO & \\
 &   &  &  & \\
ad3a-22603 & ... & 16:07:07.08 & extended CS, & O-AGB either near\\
 &  & $-$52:23:43.8 & point SiO & or displaying C-bearing\\
 &  &  &  & molecular emission\\
ad3a-26186 & ... & 17:12:08.54 & extended \& & \small{CS source is molecular cloud}\\
 &  & $-$38:30:20.5 & off-center CS & \small{near BAaDE target AGB}\\
  &  & &  & \\\
ad3a-26358 & IRAS 17144$-$3745 & 17:17:52.79 & extended CS, & O-AGB displaying emission\\
 &  & $-$37:49:07.9 & point SiO, & from C-bearing molecules\\
  &  &  & extended H$^{13}$CN & concentric to SiO\\
ad3a-26862 & IRAS 17043$-$3551 & 17:07:43.54 & point CS, & C-star with thermal SiO\\
 &  & $-$35:55:51.1  &  point H$^{13}$CO$^+$, & \\
  &  &   &  point SiO & \\
ad3a-27026 & 2MASS J17262767 & 17:26:27.67 & extended CS, & O-AGB displaying emission\\
 & -3521552 & $-$35:21:55.2 & extended H$^{13}$CO$^+$, & from C-bearing molecules\\
 &  &  & point SiO & concentric to SiO\\
\tablecaption{CS detections that lie in the O-AGB region} 
\enddata
\tablecomments{Names, BAaDE data, and suggested explanation for the 6 sources whose colors indicate they are O-rich AGB stars but whose spectra show a detection of CS. More thorough explanations are in Sect\,\ref{csNoagb}.}
\end{deluxetable*}

\subsubsection{CS detections in the O-AGB color-region}\label{csNoagb}
First, we discuss CS detections that lie in the O-AGB region (listed in Tabel 1). 5357 sources have BAaDE spectra, reliable $[K_s]$, $[A]$, and $[E]$ magnitudes, and lie in the O-AGB region of Fig.\,\ref{fig:kaae}. Of these, 3920 ($\sim$73\%) are detected in SiO, while only 6 ($\sim$0.1\%) of the O-AGB sources are detected in CS. The remaining non-detections can be attributed to variability and sensitivity (see Sect.\,\ref{detrate}). It should be noted that the number of CS detections is artificially low because only ALMA data (10\% of our sample) include spectral coverage of a CS line. Descriptions of these six sources, including their spectral content and morphology based on images in Stroh et al.\, \citeyearpar{michael2}, follow.
\begin{itemize}
    \item \textit{ad3a-22405}: We propose that this source is a candidate S-type star (defined as an AGB star with a C/O ratio close to 1). The CS emission is extended and the SiO maser emission distribution is point-like. This object also shows the v=2 $J=2-1$ line, which is rare at 86 GHz \citep{michael}. A pumping model by Olofsson et al.\,\citeyearpar{olofsson} suggests that the v=2 line is destroyed by a line overlap with H$_2$O in ordinary O-rich AGB stars, strengthening the case for S-type candidacy. Soria-Ruiz et al. \citeyearpar{stype} also observe both v=1 and v=2 lines at 86 GHz in the S-type star $\chi$ Cyg.
    \item \textit{ad3a-22603}: This source lying in the O-AGB region shows weak extended CS emission and point-like SiO maser emission. The CS emission is slightly offset from the SiO emission in both position and velocity and it is unclear whether the two are associated. This source could be a case of a CS shell or outflow, or a case of an unassociated CS-emitter near the targeted source. The issue of detecting an extended feature in the sidelobes as opposed to detecting the targeted MSX/BAaDE source may also present itself in many of the YSO region sources as they are often situated in clouds. Detections, like this one, that are not made at the phase center (but within a few arcseconds of the center) are marked in Table 6 of Stroh et al.\,\citeyearpar{michael2} and should be treated with caution.
    \item \textit{ad3a-26186}: The CS emission in this source extends across the field of view and is offset from the targeted position. This source may be detected in CS via the sidelobes simply because it is situated in a molecular cloud, where the targeted source is a non-detection (likely an O-AGB with CSE or variability conditions unsuitable for maser detection within our sensitivity). 
    \item \textit{ad3a-26358}: The SiO maser emission is point-like and contained within extended CS and H$^{13}$CN emission, in agreement with the general model where SiO masers form within the dust condensation zone \citep{habing}. This source is bright ($[A]=1.93$) and identified in the literature as an OH/IR star (OH 349.39$-$0.01; \citealt{caswell}). The source may just be an O-AGB that is sufficiently nearby to display weak signatures of C-bearing molecules, or may be more evolved and have a more detached CSE than most BAaDE sources. Most likely the distance and evolutionary status combine to make CS detectable in this source, even though the source is likely O-rich.
    \item \textit{ad3a-26862}: Displaying CS, H$^{13}$CO+, and thermal (v=0) SiO emission, all with point-like morphology, this source does not directly fit within our classification scheme for O/C-rich AGB stars, but is likely an evolved stellar source.  
    \item \textit{ad3a-27026}: The SiO v=1 detection in this source is point-like and the CS and H$^{13}$CO$^+$ emission are both extended. The source is bright ([A] = 2.02), has been identified as an O-rich AGB by Le Bertre et al.\,\citeyearpar{lebertre}, and may be another case of C-bearing molecules being detectable in an O-rich object either due to evolutionary status or proximity. 
\end{itemize}

\startlongtable
\begin{deluxetable*}{ccccc}
\tablehead{
\colhead{BAaDE name} & \colhead{Other name} & \colhead{RA and} & \colhead{BAaDE} & \colhead{Explanation}
\\ \colhead{} & \colhead{} & \colhead{Dec} & \colhead{detections} & \colhead{}
}
\startdata
ad3a-00598 & ... & 17:56:03.576 & SiO J=(1-0) & outlier O-AGB\\
 &  & $-$33:16:08.40 & v=1,2 & \\
ad3a-03104 & 2MASS J17451729$-$ & 17:45:17.232  & SiO J=(1-0) & outlier O-AGB/long-period variable\\
 & 2904133 &  $-$29:04:17.76 & v=2 & (Matsunaga et al. 2009)\\
ad3a-03262 & V4520 Sgr & 17:45:49.200 & SiO J=(1-0)  & outlier O-AGB/Mira \\
 &  & $-$28:58:45.84 & v=1,2,3, $^{29}$SiO v=0 & (Samus et al. 2017)\\
ad3a-03315 & ... & 17:45:49.392  & SiO J=(1-0) & outlier O-AGB\\
 & &  $-$28:56:52.08 & v=1,2 & \\
ad3a-06166 & OGLE BLG-LVP- & 17:41:37.080  & SiO J=(1-0) & outlier O-AGB/ Mira\\
& 17845& $-$24:24:10.08 & v=1,2 & (Soszynski et al. 2013)\\
ad3a-06182 & IRAS 17306$-$2420 & 17:33:45.480  & SiO J=(1-0) & outlier O-AGB\\
  & &  $-$24:22:06.96 & v=1,2 & \\
ad3a-06295 & ... & 17:42:45.888  & SiO J=(1-0) & outlier O-AGB\\
 & &  $-$24:04:10.56 & v=1,2 & \\
ad3a-08663 & IRAS 18132$-$1719 & 18:16:07.776  & SiO J=(1-0) & outlier O-AGB\\
 & &  $-$17:17:56.04 & v=0,1,2,3 & \\
ad3a-11594 & ... & 18:36:50.616  & SiO J=(1-0) & outlier O-AGB\\
 & &  $-$07:20:26.88 & v=1,2 & \\
ad3a-11598 & ... & 18:38:46.008 & SiO J=(1-0) & outlier O-AGB\\
 & & $-$07:20:06.00 & v=1 & \\
ad3a-13310 & IRAS 18279$-$0232 & 18:30:35.760  & SiO J=(1-0) & possible outlier YSO\\
  & & $-$02:30:37.08 & v=1,2 & (Dunham et al. 2015)\\
ad3a-22459 & IRAS 15576$-$5251 & 16:01:28.00 &  SiO J=(2-1)& outlier O-AGB\\
 & & $-$52:59:34.5 & v=1 & \\
ad3a-25994 & IRAS 17217$-$3916 & 17:25:13.07 & SiO J=(2-1) v=0, & C-AGB with thermal SiO\\
 & & $-$39:19:21.7 &  CS, H$^{13}$CN & \citep{alksnis}\\
ad3a-26120 & ... & 17:11:22.78 & SiO J=(2-1) v=0, & C-AGB with \\ 
 & & $-$38:48:16.2 &  CS, H$^{13}$CN & thermal SiO\\
ad3a-26498 & IRAS 17137$-$3713 & 17:17:07.90 & SiO J=(2-1) & outlier O-AGB\\
  & & $-$37:16:38.9 & v=0,1 & \\
ad3a-27052 & IRAS 17199$-$3512 & 17:23:18.01 & SiO J=(2-1) v=0, & C-AGB with thermal SiO\\ 
 & & $-$35:15:37.6 &  CS, H$^{13}$CN & \citep{kvb}\\
ae3a-00099 & IRAS 17464$-$2843 & 17:49:40.056 & SiO J=(1-0) & outlier O-AGB\\
 & &  $-$28:44:48.12 & v=1,2,3,$^{29}$SiO v=0 & \\
ae3a-00116 & ... & 17:47:09.168  & SiO J=(1-0) & outlier O-AGB\\
& & $-$28:35:29.04 & v=0,1,2,3 & \\
ae3a-00123 & 2MASS J17470898$-$ & 17:47:08.832 & SiO J=(1-0) v=0,1,2,3, & outlier O-AGB\\
 & 2829561 & $-$28:29:56.40 & $^{29}$SiO v=0,$^{30}$SiO v=0 & \\
ae3a-00153 & SSTM20 & 18:02:22.824 & SiO J=(1-0)& possible outlier YSO\\
 & J180222.92$-$225541.2 & $-$22:55:40.80 & v=1,2 & (Rho et al. 2006)\\
ae3a-00162 & IRAS 18002$-$2218 & 18:03:17.712 & SiO J=(1-0) & outlier O-AGB\\
 & &  $-$22:17:49.56 & v=1,2 & \\
ae3a-00171 & 2MASS J18060316$-$ & 18:06:03.288  & SiO J=(1-0) & outlier O-AGB/long-period variable\\
 & 2119472 & $-$21:19:46.20 & v=1,2,3,$^{29}$SiO v=0 & \citep{saito}\\
ae3a-00317 & ... & 18:43:59.808  & SiO J=(1-0) & outlier O-AGB\\
& & $-$03:35:46.68 & v=1,2,3 & \\
\tablecaption{SiO detections that lie in the C-AGB region}
\enddata
\tablecomments{Names, BAaDE status, and suggested explanation for the 23 sources whose colors indicate they are C-rich AGB stars but whose spectra show a detection of SiO. Most sources in this list are likely O-AGBs whose color is abnormal due to the non-simultaneity of the IR-observations, or the lack of reddening corrections (or potential misidentifications). More thorough explanations for select sources are in Sect\,\ref{sioNcagb}.}
\end{deluxetable*}

\subsubsection{SiO maser detections in the C-AGB color-region}\label{sioNcagb}
As SiO detections can be made in both the VLA and ALMA data there are more detections of SiO outside of the O-AGB region than CS within the O-AGB region. 12\% (23) of the sources that lie in the C-AGB region are SiO detections; they are listed in Table 2. Generally these sources are likely outlier O-AGBs, because they show SiO maser emission and many sources identified in the literature are Mira stars, AGB stars, or candidates thereof (see Table 2). Sources with different identifications than these are discussed as follows:
\begin{itemize}
    \item Two sources are identified as YSO candidates. Both show $[K_s]-[A]>8.7$ (i.e.\,very red) and so are in the region of Fig.\,\ref{fig:kaae} where O-AGBs, C-AGBs, and YSOs are closest to each other on the diagram. Of those two, \textit{ae3a-00153} is identified as a YSO by Rho et al. \citeyearpar{rho} through Spitzer imaging and is situated in the Trifid nebula. As our study does not correct for reddening, the IR-colors of this source could be affected by its position in the nebula. \textit{ad3a-13310} is identified by Dunham et al. \citeyearpar{dunham} and is marked in their study as not likely to be a misidentified AGB (i.e.\,the YSO designation is strongly favored in their study). These sources could be incorrectly identified, or could be YSOs with SiO maser emission; although SiO maser emission in YSOs is rare, it is known to occur in a handful of sources \citep{ysosio}.
    \item \textit{ad3a-09227} is a known S-type star \citep{nassau} which lies in the O-AGB section of Fig.\,\ref{fig:kaae} but close to the O/C boundary. Its BAaDE spectrum shows a single line detection of SiO v=1 (J=1-0) making it the only single line detection of SiO in the C-area. Using Very Long Baseline Array (VLBA) observations of $\chi$ Cyg, an S-type AGB star, to investigate emission structure around the star, Soria-Ruiz et al.\,\citeyearpar{stype} conclude that at least $\sim$80\% of the circumstellar regions emitting in the 43 GHz v = 1 line have no counterpart in the v = 2 transition in their source. The summed v = 2 flux would be much fainter than the v=1 flux in an unresolved source. This example suggests that single-line v = 1 detections in the C-AGB region may be a signature of S-type stars, but so far we have no other detections like this to propose as S-type candidates. This type of detection will be sought out as BAaDE spectra continue to be produced. 
\end{itemize}

\subsubsection{SiO maser detections in the YSO region}\label{sioNyso}
Most of the 33 sources that lie in the YSO section of Fig.\,\ref{fig:kaae} and have SiO maser emission are also associated with AGB stars and/or Mira variables, and are therefore likely outliers of the O-AGB population. All of these sources are listed in Table 3 and a couple special cases are discussed below. 
\startlongtable
\begin{deluxetable*}{ccccc}
\tablehead{
\colhead{BAaDE name} & \colhead{Other name} & \colhead{RA and} & \colhead{BAaDE} & \colhead{Explanation}
\\ \colhead{} & \colhead{} & \colhead{Dec} & \colhead{detections} & \colhead{}
}
\startdata
ad3a-00024 & IRAS & 17:50:26.184 & SiO J=(1-0) & outlier O-AGB/Mira\\
 & 17471$-$3453 & $-$34:54:27.72 & v=1,2 & (Soszynski et al.\,2013)\\ 
ad3a-00095 & IRAS &  17:21:08.136 & SiO J=(1-0) & outlier O-AGB\\
 & 17178$-$3437 & $-$34:39:54.00 & v=1,2 &\\
ad3a-00571 & 2MASS J17451164$-$ &  17:45:11.808 & SiO J=(1-0) & outlier O-AGB/Mira\\
 & 3320292 & $-$33:20:29.04 & v=1,2,$^{29}$SiO v=0 & (Soszynski 2013)\\
ad3a-00718 & IRAS 17109$-3$255 & 17:14:13.37 &  SiO J=(1-0)& outlier O-AGB\\
 & & $-$32:59:04.0& v=1,2 & \\
ad3a-00934 & ... & 17:28:50.280 & SiO J=(1-0) & outlier O-AGB\\
 & &   $-$32:22:34.32 & v=1,2 & \\ 
ad3a-01092 & ... & 17:35:45.83 & SiO J=(1-0)& outlier OAGB\\
 & & -31:54:45.0  & v=1,2 &\\
ad3a-01321 & ... & 17:46:14.280  & SiO J=(1-0)& outlier O-AGB\\
 & &  $-$31:16:44.76  & v=1,2 &\\
ad3a-01431 & IRAS & 17:31:40.344   & SiO J=(1-0)& outlier O-AGB\\
 & 17284$-$3058 &   $-$31:00:15.48  & v1,=2 &\\
ad3a-02098 & IRAS & 17:11:32.400  & SiO J=(1-0) & outlier O-AGB\\
 & 17083$-$2949 & $-$29:53:26.52 & v=1,2,3 & \\
ad3a-02480 & 2MASS J17581200$-$ &  17:58:12.000 & SiO J=(1-0) & outlier O-AGB/double star\\
 & 2929127 & $-$29:29:13.20 & v=1,2 & (Gaia)\\
ad3a-02723 & 2MASS J17524959$-$ &  17:52:49.584 & SiO J=(1-0) & outlier O-AGB/Mira\\
 & 2918426 & $-$29:18:42.48 & v=1,2 & (Soszynski et al.\,2013)\\
ad3a-02794 & V4490 Sgr &  17:45:16.416 & SiO J=(1-0) & outlier O-AGB/ OH/IR\\
 &  &  $-$29:15:39.96 & v=1,2 & (Lindqvist et al.\,1992)\\
ad3a-03102 & V4953 Sgr &  17:45:50.784 & SiO J=(1-0) & outlier O-AGB/Mira\\
 &  & $-$29:04:21.00 & v=1,2 & (Samus et al.\,2017)\\
ad3a-03166 & ... &  17:45:38.640 & SiO J=(1-0) & outlier O-AGB\\
 &  & $-$29:02:06.72 & v=1,2,3 &\\
ad3a-03210 & 2MASS J17493768$-$ & 17:49:37.512 & SiO J=(1-0) & outlier O-AGB \\
 & 2900476 & $-$29:00:49.32 & v=1,2 & \\ 
ad3a-03247 & V5002 Sgr &  17:46:06.408 & SiO J=(1-0) & outlier O-AGB/Mira\\
 & & $-$28:59:08.52 & v=1,2,$^{29}$SiOv=0 & (Samus et al.\,2017)\\
ad3a-03311 & V5085 Sgr & 17:46:30.408 & SiO J=(1-0) & outlier O-AGB/LPV\\
 &  &  $-$28:56:58.20 & v=1,2 & (Glass et al.\,2001)\\ 
ad3a-03396 & V4985 Sgr & 17:46:02.184 & SiO J=(1-0) & outlier O-AGB/LPV\\
 & & $-$28:53:15.36 & v=1,2,3 & (Glass et al.\,2001)\\
ad3a-03423 & ... & 17:48:40.944  & SiO J=(1-0)& outlier O-AGB\\
 & &  $-$28:52:14.52  & v=2 &\\
ad3a-03436 & ... & 17:48:36.336  & SiO J=(1-0) & outlier O-AGB or \\
 &  & $-$28:51:47.52.408 & v=1 & cross match error\\
ad3a-03628 & 2MASS J17461885$-$ &  17:46:18.720 & SiO J=(1-0) & outlier O-AGB/LPV \\
 & 2844392 & $-$28:44:39.12 & v=1,2 & (Matsunaga et al.\,2009) \\
ad3a-03686 & ... & 17:48:29.688 & SiO J=(1-0) & outlier O-AGB\\
 & & $-$28:42:53.28 & v=1,2 &\\ 
ad3a-04003 & 2MASS J17465614$-$ & 17:46:55.896 & SiO J=(1-0) & outlier O-AGB\\
 & 2831026 & $-$28:31:03.00 & v=1,2 &\\
ad3a-04066 & ... & 17:47:16.848 & SiO J=(1-0) & outlier O-AGB\\
 &  & $-$28:27:50.04 & v=1,2 &\\
ad3a-04466 & IRAS & 18:10:38.736 & SiO J=(1-0) & outlier O-AGB\\
 & 18074$-$2805 &  $-$28:05:04.56 & v=1,2 & \\
ad3a-06511 & ... & 18:00:15.600 & SiO J=(1-0) & outlier O-AGB\\
 & &  $-$23:28:54.48 & v=1,2 &\\ 
ad3a-07392 & IRAS &  18:19:20.064 & SiO J=(1-0) & outlier O-AGB\\
 & 18163$-$2151 &  $-$21:49:49.08 & v=1,2,$^{29}$SiO v=0 &\\
ad3a-08123 & ... & 17:53:35.568  & SiO J=(1-0) & outlier O-AGB\\
 & &  $-$19:18:01.80 & v=1,2 & \\
ad3a-08677 & IRAS & 17:59:33.336 & SiO J=(1-0) & outlier O-AGB \\
 & 17566$-$1713 & $-$17:13:28.92 & v=1,2 & \\ 
ad3a-09754 & ... &  18:25:21.456 & SiO J=(1-0) & outlier O-AGB\\
 &  & -13:08:56.76 & v=1,2 & \\
ad3a-26529 & ... & 17:19:52.21 & SiO J=(2-1) v=0, & outlier C-AGB\\
 & & $-$37:09:28.1 & CS, H$^{13}$CN &\\
ad3a-26782 & ... &  17:19:41.00 & SiO J=(2-1) & outlier O-AGB\\
 &  & $-$36:10:19.6 & v=1 &\\
 ad3a-26826 &... & 17:19:47.87 & SiO J=(2-1) & outlier O-AGB\\
 & & $-$36:03:23.0 & v=1 &\\
\tablecaption{SiO detections that lie in the YSO region}
\enddata
\tablecomments{Names, BAaDE data, and suggested explanation for the 33 sources whose colors indicate they are YSOs but whose spectra show a detection of SiO. Most sources in this list are likely O-AGBs whose color is abnormal due to the non-simultaneity of the IR-observations, or the lack of reddening corrections (or potential misidentifications). More thorough explanations are in Sect\,\ref{sioNyso}.}
\end{deluxetable*}

\begin{itemize}
    \item The star VX Sgr is BAaDE source \textit{ce3a-00110}, which shows multiple SiO transition detections ($^{28}$SiO v = 0, 1, and 2, $^{29}$SiO v= 0, and $^{30}$SiO v= 0) and is a known red supergiant 1.56 kpc from the Sun \citep{xu}. Whereas it is rightfully classified as an O-rich star by our cuts, it lies close to the border between O-AGBs and non-AGB objects. 
    
    \item The source \textit{ad3a-26529} shows the traits indicative of a C-AGB discussed in Sect.\,\ref{crichsect} although its color is very unusual for an AGB star (very red in $[A]-[E]$ and $[D]-[E]$). As its spectrum only shows thermal SiO emission, it is not too unusual for it to lie outside of the O-AGB region, but its status beyond non-O-AGB is undetermined. 
\end{itemize}   
\par Many SiO detections within the non-AGB section are not yet associated with a source of known type. We suggest that these unassociated sources, especially ones that lie within or near molecular clouds/star-forming regions, could be YSOs with SiO maser emission or other super-giants as both types of sources are dynamically young and are expected to be near the clouds they formed from. Other SiO detections in this area could belong to post-AGB objects as post-AGB objects and YSOs share similar IR colors in many bands.

\section{Discussion}\label{discussionsect}
With the IR-based divisions between the populations in our sample established, we first discuss possible reasons behind the color-differences (or lack thereof) between populations (Sect.\,\ref{emag}, \ref{ocindexsect}). These divisions tightly constrain the regions of the $[K_s]-[A]$ versus $[A]-[E]$ diagram where we expect to find SiO masers, and we  next discuss the impact of this on the BAaDE detection rates (Sect.\,\ref{detrate}). Finally sorting our sources by colors, we discuss the O-AGB to C-AGB ratio and the distributions of each population in the Milky Way (Sect.\,\ref{numbersect}, \ref{distributionsect}). Detailed modeling and discussion of the corresponding  shell chemistry and composition yielding these IR colors is beyond the scope of this work.

\subsection{Significance of [E]-magnitude} \label{emag}
\par $[E]$-magnitude information is critical in identifying the YSO population. The AGB and YSO populations are not separable in MSX colors that do not utilize $[E]$-magnitude, including $[A]-[D]$ (the primary color used for BAaDE selection), which is why these sources were not deselected initially. Relying more heavily on selection criteria from SCC09 involving the [E]-magnitude (picking a greater percentage of sources based on [A]$-$[E] or [C]$-$[E] as opposed to [A]$-$[D] colors) could have excluded most of these sources. 

\par Unfortunately, the $[E]$-band sensitivity is lower than that of the other MSX bands ($[A]$-band is 10 times more sensitive), and only $\sim45\%$ ($\sim$ 6000) of our sample have reliable (quality 3 and higher) $[E]$-magnitudes. This means that even though the distinction between AGB stars and other sources is clear, we can only definitively separate these two groups for about half of our sample. 

\subsubsection{Alternatives to [E] and the 18-micron feature}
\begin{figure}
    \centering
    \includegraphics[width=.47\textwidth]{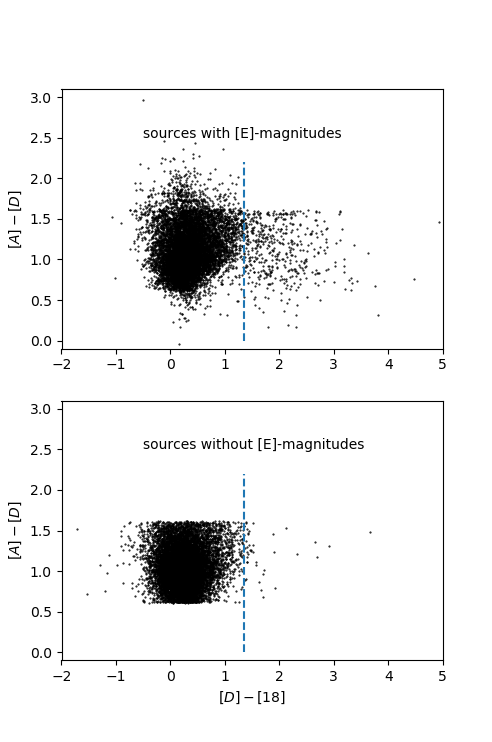}
    \caption{Zero-magnitude-corrected color-color plots of sources from our sample with (top panel) and without (bottom panel) reliable [E]-magnitudes. All sources in both panels have been cross-matched with AKARI and have reliable [18]-magnitudes. The dashed line indicates the value of our [D]$-$[E] cut that divides AGBs and YSO. It is clear that [D]$-$[18] also shows a distinct red population similar to the DE-red sources. Sources without reliable [E]-magnitudes are almost exclusively found on the left-hand side of the dashed line (i.e., preferentially blue in [D]$-$[18]). The cuts in [A]$-$[D] color are less distinct in the top panel than in the bottom because sources were selected on [A]$-$[D], [C]$-$[E], or [A]$-$[E] color in the top panel and only on [A]$-$[D] in the bottom panel because of the lack of reliable [E]-magnitudes.}
    \label{d18ad}
\end{figure}

Because many BAaDE sources lack reliable [E]-magnitudes (centered at 21 $\mu$m), we here explore IR-band options from other surveys that may separate YSO and AGB sources. The cross-matching between these surveys is discussed in Sect.\,\ref{irselectsect}. Both [WISE4] and MIPSGAL [24] (centered at 22 and 24 microns respectively) are similar in center wavelength and bandwidth to MSX [E]; however when applied to our [D]$-$[E] selection, neither [D]$-$[WISE4] nor [D]$-$[24] show significant separation of a red-colored low-detection-rate population from the bulk of the sources. In contrast, although AKARI [18] (centered at 18$\mu$m) is much wider than [E] and is centered at a shorter wavelength, [D]$-$[18] does show a significant red-colored low detection-rate population that closely matches the MSX-defined DE-red population (see top panel of Fig.\,\ref{d18ad}). Although [WISE4], MIPSGAL [24], and [E] are all similar, [E] and AKARI [18] are the only two of the four bands that cover the 18 $\mu$m silicate feature; the other two filters fall off shortward of around 20 $\mu$m. This indicates that the $18\mu$m silicate feature likely determines the DE-red/DE-blue and YSO/AGB division \footnote{AGB sources in IRAS region IIIa (and thus MSX iiia) have been characterized as having the 18$\mu$m feature in emission, making for brighter [E] and [18]-magnitudes.}. 
\par As only 47\% of the complete BAaDE sample has reliable [E]-magnitudes, using AKARI [18] as well nearly doubles the number of sources in our sample that can be classified. However, although a significant difference between red and blue populations can be seen, the actual division at a specific color value is less evident in [D]$-$[18] than in the original DE separation, and therefore drawing a line to separate the two is more straightforward in [D]$-$[E]. Additionally, sources that have [18]-magnitudes but no [E]-magnitudes are strongly preferentially [D]$-$[18] blue and so appear to be AGB stars based on [D]$-$[18] color (see Fig.\,\ref{d18ad}). Separating our sample by sources with or without reliable [E] data, we find [18]-magnitudes are typically clearly 1-2 magnitudes fainter for the sources without [E] (in the range $1<[18]<4.5$
versus $-1<[18]<3.5$ mag for sources with [E]). It seems that most sources that are not detected in [E] are not detected because their [E]-magnitude emission is weak compared to their [D]-magnitude emission, and that these sources are therefore not likely to be YSOs as defined by our color-cut. We therefore define all DE-blue \emph{and all} sources without [E]-magnitudes as likely AGB stars, and only exclude DE-red sources from the AGB population. 

\begin{figure*}[tb]
    \centering
    \includegraphics[width=.9\textwidth]{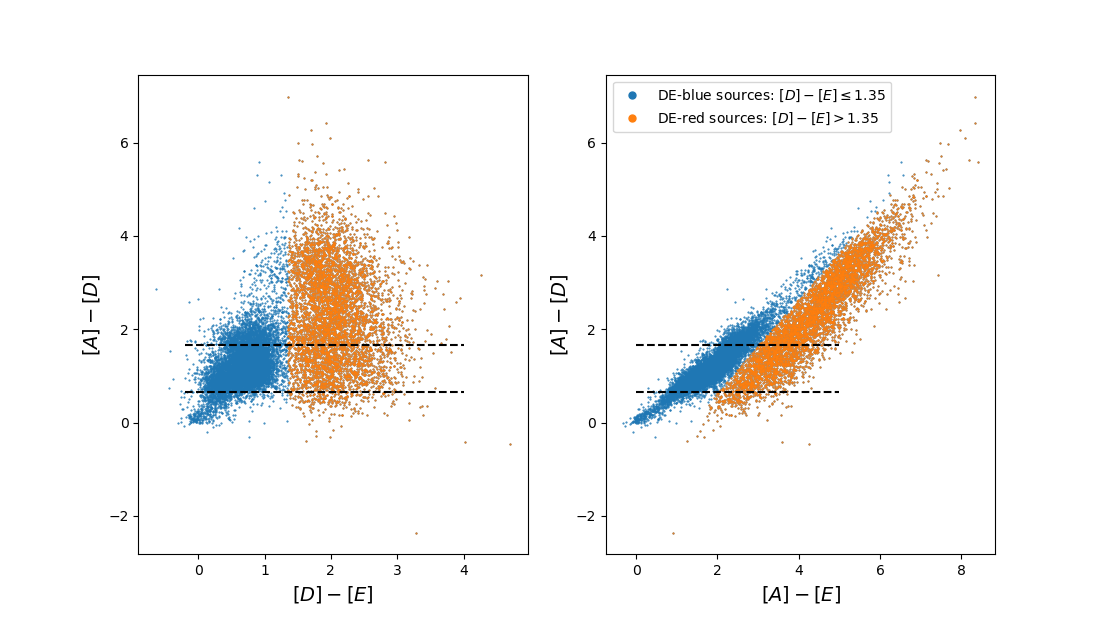}
    \caption{Zero-magnitude-corrected color-color plots of all 15219 MSX sources with $[E]$-magnitudes. The DE-red sources within the currently available BAaDE data show a very low rate of SiO detections (see Fig.\,\ref{fig:dead}), and we conclude they are not AGB stars but most likely YSOs. We can, however, only make a determination about the properties of sources within the BAaDE $[A]-[D]$ color range enclosed by the dashed lines, as this is how 97\% of the BAaDE sources were selected. }%
    \label{fig:allmsx}
\end{figure*}
\subsubsection{Extension to the entire MSX PSC}
The DE-red population is easily identifiable in the entire MSX sample (Fig.\,\ref{fig:allmsx}), not only in the BAaDE subset, and was seen as an offset population in an $[A]-[D]$ versus $[A]-[E]$ diagram by SCC09. SCC09 suggested that the offset might have been caused by an error in the $[E]$-band photometry, but we find that, because the SiO detection rate is strongly linked to whether a source is DE-red or not, the offset likely has a physical basis. As we do not have SiO/CS emission information of MSX sources outside of $0.66 < [A]-[D] < 1.66$ we cannot confirm that \emph{all} MSX sources with $[D]-[E]>1.35$ are related to a YSO population. A study with a less limited $[A]-[D]$ color range is needed to determine how the BAaDE DE-red population relates to the entire MSX sample.
\subsection{Revisiting the OC-index} \label{ocindexsect}

\begin{figure}
    \centering
    \includegraphics[width=.47\textwidth]{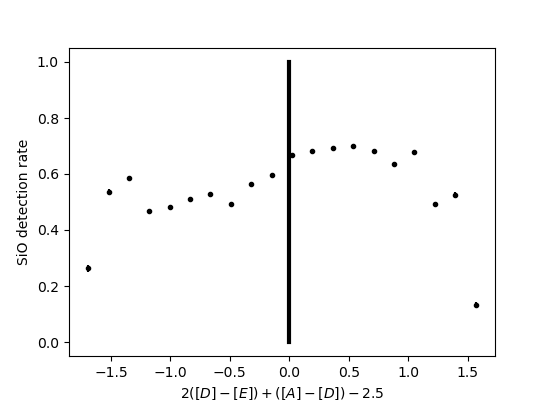}
    \caption{BAaDE SiO maser detection rate per 0.17 magnitude bin as a function of the OC-index defined in SCC09. DE-red sources and bins with fewer than 10 sources have been removed. Error bars are often smaller than the symbol size. Previous surveys have used this index as a predictor of whether a source is O- or C-rich (see Sect.\,\ref{ocindexsect}), where a negative index predicts a more likely C-rich source and positive predicts more likely C-rich; the solid line shows the division between the two. Our sample only shows a slight decrease in detection rate for negative indices, therefore OC-index is not a reliable predictor of AGB chemistry in our sample.}
    \label{ocindex_old}
\end{figure}

The O- and C-rich populations are not clearly separable in MSX colors alone. This was a major conclusion of SCC09 and can now be seen through the consistency of the SiO detection rate throughout the AGB sample on MSX color-color diagrams. For example, there is a lack of a clear division even in the $[A]-[D]$ versus $[D]-[E]$ diagram in Fig.\,\ref{fig:dead}, despite the fact that this color-color diagram has shown a clear distinction between O- and C- stars in smaller samples in the literature \citep{ortiz, lumsden, suh}. SCC09 define an `OC-index' in their Fig.\,5 as the distance from the OC-dividing line (shown as the diagonal line in Fig.\,\ref{fig:dead}), which is a predictor of whether a source from the Ortiz et al.\,(2005) sample will be an O- or C-rich AGB. Negative indices indicate a more likely C-star and positive indices indicate more likely O-rich AGB stars. SCC09 found that the OC-index is not a secure predictor of whether cross-matched MSX and IRAS sources appear in a C- or O- dominated region of the IRAS diagram, but that the index is more suited to divide stars based on their temperature. Indeed, in the SCC09 study many IRAS-defined O-AGBs are found on the C-side of the OC-dividing line. Conversely, OC-index defined C-stars include most sources with $[12]-[25]\leq-1.5$, not only those in regions VIb and VII (designated for C-stars), showing that the OC-index is more closely related to $[12]-[25]$ color than to source chemistry (see SCC09 Sect.\,3.2 for details). 
\par In Fig.\,\ref{ocindex_old} the SiO detection rate is relatively consistent across OC-index with only a small decrease towards the more likely C-rich stars, especially as compared to Fig.\,\ref{ocindex_new}. Therefore, in our sample, the OC-dividing line is ineffective; we find that the OC-index is not a reliable predictor of AGB chemistry within the BAaDE survey, which is in agreement with SCC09. 
\par Our primary explanation as to why this separation is so distinct in other studies but lacking in our sample lies in the way the samples were collected. The BAaDE sample is pseudo-blind when it comes to O/C-AGBs (although we may cut out proportionately more C-AGBs than O-AGBs with our [A]$-$[D]$>0.66$ cut, see Sect.\,\ref{numbersect}). The Lumsden et al.\,(2002), Ortiz et al.\,(2005), and Suh et al.\,(2011) samples were collected from the literature. Lumsden et al.\,(2002) uses a compilation from Alksnis et al.\,(2001) as their identified C-stars. These sources were mostly optically selected and are therefore likely biased towards thin-shell, bluer C-stars. For identified O-rich AGB stars, Lumsden et al.\,\citeyearpar{lumsden} use the OH/IR stars from Chengalur et al. (1993), which biases their O-stars towards the thicker-shell redder OH/IR population. Ortiz et al.\,\citeyearpar{ortiz} suffer similar biases; their carbon stars come from the Loup et al.\,\citeyearpar{loup} sample which is biased towards strong CO emitters and includes mainly sources for which $[12]-[25]<-0.3$ (see Loup et al. Fig.\,3), and their O-stars are again primarily OH/IR sources. In short, these samples have selected thin-shell, blue-biased carbon sources, and thick-shell (OH/IR) red-biased oxygen sources, which has exaggerated the distinction between the two groups in MSX color-color space. Suh et al.\,\citeyearpar{suh} include some SiO-detected O-AGBs in their sample, which lessens this bias by including more thin-shell O-AGBs (as the thinner shells are more conducive to SiO masers), and indeed they do see cross-over of O-rich objects into the C regime, although their sample still separates reasonably well along the OC-dividing line. BAaDE focuses strongly on thin-shell O-AGBs, as this is where SiO emission is most common, and the sample shows almost no separation by the OC-index. As such we conclude that the OC-index is not suitable for separating thin-shell O-AGBs from C-AGBs. 
\par Within the BAaDE sample, a better division between O- and C-AGBs is found when including [K$_s$]-band measurements. Comparing Figs.\,\ref{fig:kaae} and \ref{fig:dead} shows the strong improvement that including [K$_s$]-band measurements makes on the dependability of color-based predictors of SiO maser activity, and therefore AGB chemistry.
\subsection{Detection rates within the O-AGB region}\label{detrate}
With a tightly-defined region of our color-color diagram designated for O-AGBs from Fig.\,\ref{fig:kaae} the effects of a) sensitivity and variability, b) CSE conditions, and c) non-O-AGB objects in the O-AGB section on our detection rate can be better constrained.
\par
\begin{figure}[htb]
    \centering
    \includegraphics[width=.45\textwidth]{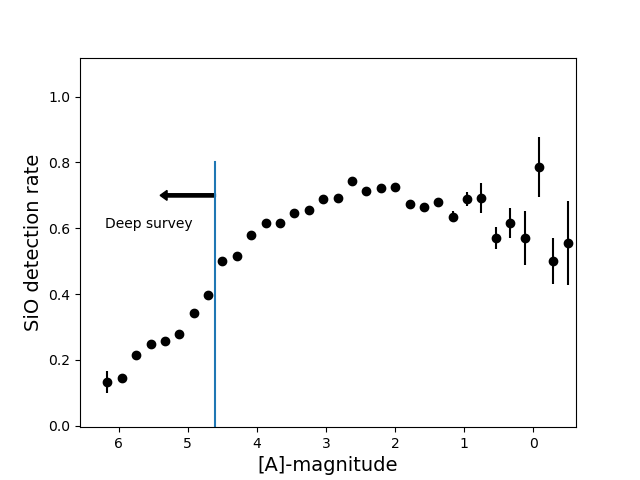}
    \caption{SiO maser detection rates per 0.2 magnitude bin as a function of $[A]$-magnitude for our sample of BAaDE sources. Error bars are often smaller than the symbol size. The detection rate drops as the $[A]$-magnitude dims, which is consistent with both magnitude and detection rate being dependent on distance. The MSX survey did not observe to even depths at all longitudes and only the deeper scans contain sources at [A]-magnitudes of 4.6 and dimmer.}
    \label{detrateA}
\end{figure}

 Sources that host consistently weak masers throughout the stellar cycle, or masers that temporarily were very weak at the time of the BAaDE observations may be non-detections in the BAaDE survey. It is likely that such a population exists, which would be part of the O-AGB region non-detections. In the O-AGB region of Fig.\,\ref{fig:kaae}, the SiO detection rate is 73\%. This rate depends on the MSX [A]-magnitude (Fig.\,\ref{detrateA}), with a peak rate around [A] $\approx 2.5$. The sources with magnitudes of 4.6 and fainter should be treated separately; these weaker sources are found in specific, deeper raster scans performed by MSX in select regions of the plane \citep{msxpsc}. Thus, in some regions the BAaDE sample contains a set of fainter sources in addition to the typical ones. A lower SiO detection rate in those regions may be expected due to sensitivity if most of these sources are fainter because they are farther from the Sun. We conclude that sensitivity negatively affects the detection rate of these fainter sources. Removing the fainter sources provides a uniformly estimated SiO detection rate of 75\% for the O-rich AGB stars with [A] magnitudes brighter than 4.6 (the main part of our sample). 

\par To address O-rich sources that do not host SiO masers at any time we rely on IR data again as IR data reflects the conditions in the CSE. If we exclude sources with $[K_s]-[A]<1.5$ (which show a lower detection rate, see Fig.\,\ref{fig:kaae}) our detection rate jumps to 80\%.  This could mean that O-AGB region sources with $[K_s]-[A]<1.5$ are O-AGBs with CSE conditions less favorable to maser conditions (perhaps too thin-shelled), while their $[K_s]-[A]\geq1.5$ counterparts are almost exclusively SiO maser sources that turn on and off as a function of stellar cycle. 
\par For comparison, Stroh et al.\,\citeyearpar{michael} report a 78\% detection rate (66/86) when re-observing a sample of known, bright SiO maser sources from the BAaDE survey. This indicates that even in a sample of previously detected sources, a $\sim$78\% detection rate is expected due to variability effects. Similarly, Pihlstr\"om et al.\,\citeyearpar{ylva} redetected 26 SiO masers out of a sample of 33 BAaDE sources which had previously shown emission, again showing that a detection rate slightly below 80\% is to be expected. As such, we conclude that variability and weak masers below our sensitivity may be responsible for most of the non-detections reported in our O-AGB sample and that, at most, only a small percentage of non-detections can be attributed to sources that are contaminants in our color-regions. In short we postulate that nearly all BAaDE selected O-AGB color-region sources are SiO emitters, but that some are just in a part of their stellar phase where their masers are off or below our moderate sensitivity.
\par O-AGBs were the intended targets of the BAaDE survey, as their maser emission provides line-of-sight velocities with which to study Galactic dynamics. Of the 28062 sources covered by the entire BAaDE survey only 2335 (8\%) do not fall into the O-AGB region based on their $[D]-[E]$, $[A]-[E]$, and $[K_s]-[A]$ colors. All other BAaDE sources either fall into this region (10084 sources), or do not have reliable data for one or more of these colors (15643 sources). 
Most of the sources which cannot be categorized based on these colors are missing [E]-magnitude information, but can be categorized as likely AGB stars based on [D] $-$ [18]. 
The original selection of sources for the survey was therefore very effective and the dynamics goal of BAaDE is not compromised by the small number of candidate C-AGBs and non-AGBs in the sample. At our current detection rate we expect $\sim$14-18 thousand SiO maser detections from the survey.

\subsection{Numbers of C versus O}\label{numbersect}
\begin{figure*}[tb]
    \centering
    \includegraphics[width=.98\textwidth]{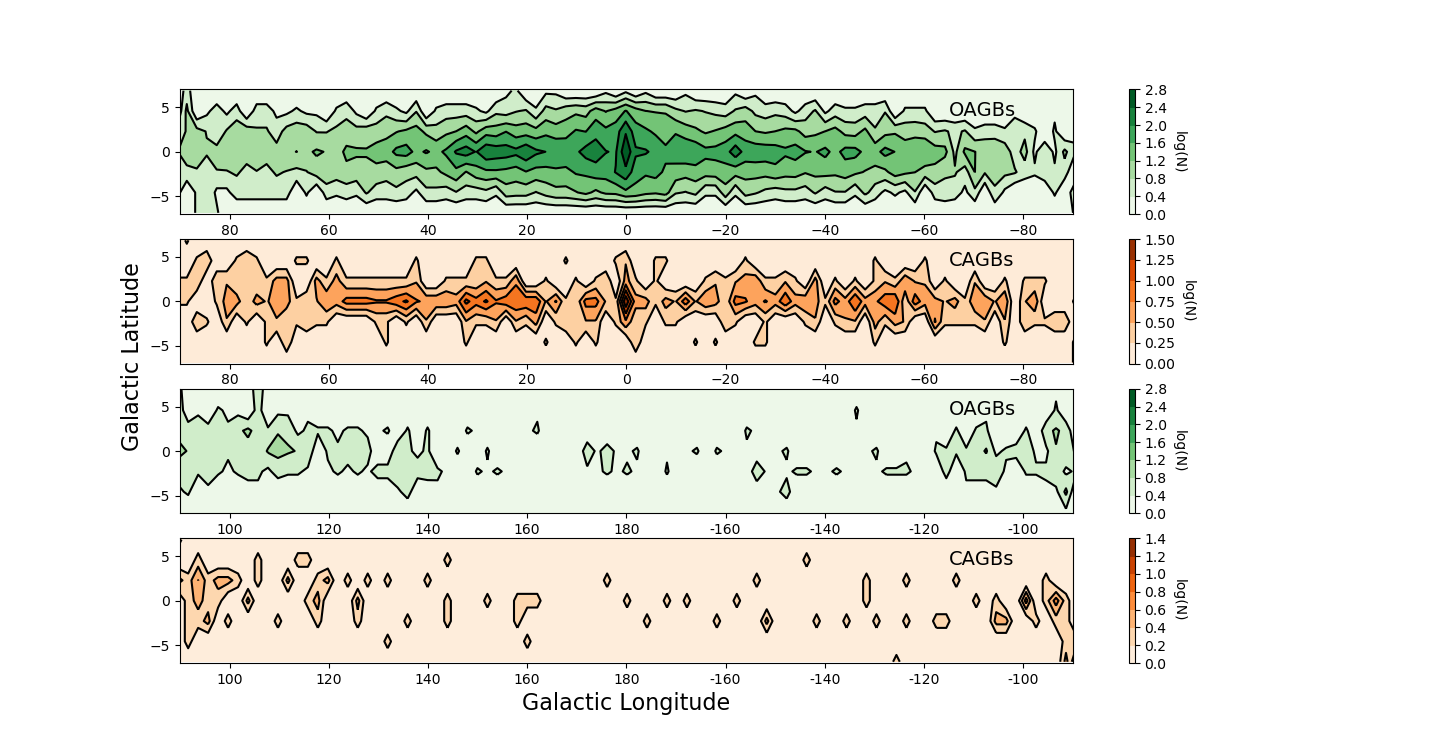}
    \caption{Distribution of O-AGBs and candidate C-AGBs in the Galactic plane. AGB stars are taken from the entire BAaDE sample as sources with [D]$-$[E]$<$1.35 and then divided into C- and O-rich groups based on Eq.\,1. The number density in each 2$^{\circ}$ bin is scaled logarithmically. O-AGBs are very concentrated near the central longitudes, while C-AGBs are more uniformly distributed.}
    \label{galplane}
\end{figure*}

Our C/O division was constructed by examining BAaDE detections and non-detections but ultimately only relies on IR colors. We can therefore apply the C/O division to all of the sources in the BAaDE sample regardless of whether they have been observed and/or analyzed yet. Within the entire BAaDE sample, of the 13120 sources that have both reliable $[E]$ and $[K_s]$-magnitudes, we find that our color criteria would yield 697 C-AGBs, 10084 O-AGBs, and 1638 YSO/non-AGB sources. This gives a fraction of C-AGBs of 6\%, but this fraction is both approximate and only applicable to the color and latitude ranges of the BAaDE survey. Firstly, our division between O- and C-AGBs is empirical and slight shifts of the division can alter this fraction by several percentage points. Secondly, a small percentage of non-detections in the O-AGB or YSO regions may belong to C-AGBs whose derived colors place them in the wrong region of our diagrams. Additionally, although we find that the OC-index is not a useful predictor of chemistry within our color-selection, it is possible that outside of our color selection, specifically towards lower (bluer) $[A]-[D]$ values, the fraction of C-AGBs could increase, as would be consistent with Ortiz et al. \citeyearpar{ortiz}. Therefore our C/O fraction is not only approximate but also only representative of AGB stars satisfying $0.66<[A]-[D]<1.66$, and other surveys with wider color-coverage are required to fully understand the effect introduced by the BAaDE color-selection. Using BAaDE sources to determine this ratio also introduces spatial effects; the BAaDE sample is representative of sources which lie close to the Plane, most of which are likely to lie in the Galactic bulge.
\par Regardless uncertainties and the limits of our sample, we can compare our fractional distributions among source types to several other surveys in the Galaxy in order to start to quantify the ratio of C-AGBs. Here we highlight studies that use differing methods for chemical identification. Kwok, Volk, \& Bidelman \citeyearpar{kvb} categorize 11224 IRAS Low-Resolution Spectra (LRS) and compare many of their classifications to those in the literature. Based on the literature identifications collected in their work, they find that 13\% (or 571 out of 4277) of AGB stars are C-rich. The identifications by Kwok, Volk, \& Bidelman \citeyearpar{kvb} based on IRAS LRS show 715 C-type and 3931 combined A- and E-type spectra (where A- and E-type sources show the 9.7 $\mu$m silicate feature in absorption and emission, respectively---both indicative of O-rich AGB stars). Excluding F-type, featureless, spectra which have more ambiguous chemical types, this puts the total number of identified AGB stars at 4646 and the fraction of C-AGBs at 15\% (715/4646). Le Bertre et al. \citeyearpar{lebertre} find that using O- and C- identifications from the InfraRed Telescope in Space (IRTS) LRS yields 18\% (126/689) C-AGBs and find that their biases should equally effect C- and O-AGBs. A study by Ishihara et al. \citeyearpar{ishi} which uses IR color cuts to define the chemical types of AGB stars (similar to this study) finds 33\% (5537/16953) of their sample to be C-rich, but puts their completeness for C-AGBs at more than double their completeness for O-AGBs (64\% and 29\% respectively) suggesting that the percentage of C-AGBs could be about 2 times too high. By these comparisons our estimated percentage of C-AGBs is relativity low, and may imply that the number of carbon stars increases at lower values of $[A]-[D]$. Physically we may also expect our sample to yield a low number of C-AGBs because we select mainly for thin-shelled sources. This means that many of our sources may be too early in their evolution or too low-mass to be C-AGBs. Furthermore, the BAaDE ratio is dominated by sources in the plane (Fig.\,\ref{lathist}) while some of the ratios listed are based on all-sky surveys or surveys that significantly cover areas away from the plane (IRAS and IRTS) and the C/O-AGB ratio may be lower in the metal-rich plane than at higher latitudes. Finally, BAaDE likely includes many Bulge sources which may also contribute to the low ratio of C-AGBs, as the Bulge is known to include mostly O-rich AGB stars. We find a 6\% fraction of C-AGBs reasonable for our Bulge dominated population.

\subsection{C- and O-AGB spatial distribution}\label{distributionsect}

\par Comparing the distributions in the Galactic plane of each of the populations as found in the entire BAaDE sample generally supports the identification of each population (Fig.\,\ref{galplane}). As seen in Sect.\,\ref{ysosect}, the BAaDE identified YSOs are found primarily close to the plane (Fig.\,\ref{lathist}), their distribution is  more clumped than the AGB stars, and shows a tendency towards molecular clouds (Fig.\,\ref{ysoclump}), all of which is consistent with YSOs. 

\par Multiple studies have shown that C-stars are more uniformly distributed across the plane than their O-rich counterparts which are preferentially found near central longitudes (e.g., Ishihara et al.\,2011, Noguchi et al.\,2004 and references therein). Our sample also shows this trend. Fig.\,\ref{galplane} shows that BAaDE color-selected O-AGB sources show a strong tendency towards the center of the Galaxy while their C-counterparts span a large range of longitudes at a fairly constant number density. The dependence of the C/O ratio on longitude is likely a metallicity effect \citep{metal}.

\subsubsection{C-AGBs in the Bulge}
Although the fraction of candidate C-AGBs relative to their O-rich counterparts drops significantly close to $l=0^{\circ}$, there are still 228 color-selected C-AGBs at $|l|<30^{\circ}$. This number is quite large given the reported paucity of C-stars in the Bulge, especially C-Miras \citep{matsunaga}. Some of the BAaDE C sources near central longitudes may be attributed to foreground objects that are closer to the Sun than typical Bulge sources, but based on source magnitudes it is unlikely that many are foreground stars. Trapp et al.\,\citeyearpar{trapp} make a first-order distinction between Bulge and Disk sources within BAaDE based on [K$_s$]-magnitude. Sources with high velocity dispersion typically have [K$_s$] values fainter than 6.0 while those with low velocity dispersion have [K$_s$] values brighter than 6.0 with a sharp distinction between the two groups. Thus sources with [K$_s$] $>$ 6.0 are likely in the Bulge. By this account 200 (88\%) of the C-AGBs within $|l|<30^{\circ}$ are likely to lie in the Bulge based on their [K$_s$]-magnitude. This means that not many of our C-AGB stars can be attributed to foreground objects even at longitudes associated with the Bulge. 
\\There could be several reasons for the large number of color and magnitude identified Bulge candidate C-AGBs. The aforementioned Trapp et al.\,\citeyearpar{trapp} result may not hold for C-AGBs as their results were based on a smaller sample from the BAaDE survey which was dominated by O-AGBs, and therefore their cut may only hold for O-AGB stars. If [K$_s$]-magnitudes cannot be used to separate Disk and Bulge C-AGBs then a large fraction of our color-identified candidate C-AGBs could be foreground objects. However, it is likely we have identified a large number of candidate C-AGBs in the Bulge with BAaDE. Our identification strategy relies on IR and radio data that, due to the affects of extinction, is more sensitive in the Bulge than obtaining optical/IR spectra, for example. Therefore, our strategy may have identified a larger number of the C-AGBs in the Bulge than was previously possible. 

\section{Conclusion}\label{conclusion}

\par We use the BAaDE SiO maser survey to study AGB star chemistry throughout the Galaxy. The survey proves an efficient tool because SiO masers likely trace O-rich AGB stars. We are able to test several IR-color division schemes of O- and C-rich AGB stars by following IR trends in SiO detection rate, and we find a clear division between potential O-rich and C-rich stars in $[K_s]-[A]$ versus $[A]-[E]$ color-color diagrams. With this distinction we are able to tightly define the region of color-color space where we expect O-AGBs to reside. The SiO detection rate within our population of O-AGBs is 73\%, which is consistent with most of the non-detections being caused by only stellar variability. 
\par We find no C/O division in any color-color diagram that utilizes only MSX data. The initial IR selection of the BAaDE sample, which includes only sources from MSX region \textit{iiia}, dictates that our sample includes many thin-shell O-rich objects which hinders finding such a division in purely MSX colors. 
\par Our sample contains a small population of red YSO sources, which was not expected based on the initial BAaDE IR color selection. This small population is easily distinguishable in MSX $[D]-[E]$ color, can be easily removed, and therefore does not contaminate the AGB and dynamical science that BAaDE is designed for. $[D]-[18]$ can also be used, to a lesser extent, to separate these populations. 
\par The distribution of the three populations (O-AGBs, C-AGBs, and YSOs) in Galactic coordinates matches expectations, with YSOs residing nearest the plane mainly in clumps in and around molecular clouds, and the O-rich AGB stars being more concentrated towards $l=0$ than the C-rich AGB stars. We find that, based on IR-color indicators, candidate C-rich AGB stars make up $\sim6\%$ of our AGB population, which is lower than many other surveys and studies. However, this fraction is approximate and may be the product of the narrow color-range used in our sample. In addition, the galactic distribution of the C-AGB/O-AGB ratio also contributes to our low value as our sample is dominated by Bulge sources with a traditionally low C-abundance.

\par Our division scheme can be used on large scales to prepare other surveys and also identify many potentially interesting targets, including S-type stars and candidate YSOs with SiO masers. A set of candidate C-rich AGB stars has been identified within the BAaDE sample, which if confirmed, might imply a larger number of C-rich stars in the Bulge region than previously thought.

\acknowledgments
Support for this work was provided by the NSF through the Grote Reber Fellowship Program administered by Associated Universities, Inc./National Radio Astronomy Observatory. The National Radio Astronomy Observatory is a facility of the National Science Foundation operated under cooperative agreement by Associated Universities, Inc. The BAaDE project is funded by National Science Foundation Grant 1517970 to UNM  and 1518271 to UCLA. This paper makes use of the following ALMA data: ADS/JAO.ALMA\#2013.1.01180.S, ADS/JAO.ALMA\#2015.1.01289.S, and ADS/JAO.ALMA\#2017.1.01077.S. ALMA is a partnership of ESO (representing its member states), NSF (USA) and NINS (Japan), together with NRC (Canada), MOST and ASIAA (Taiwan), and KASI (Republic of Korea), in cooperation with the Republic of Chile. The Joint ALMA Observatory is operated by ESO, AUI/NRAO and NAOJ. This research made use of the SIMBAD database, operated at CDS, Strasbourg, France. This research made use of data products from the Midcourse Space Experiment. Processing of the data was funded by the Ballistic Missile Defense Organization with additional support from NASA Office of Space Science. This publication makes use of data products from the Two Micron All Sky Survey, which is a joint project of the University of Massachusetts and the Infrared Processing and Analysis Center/California Institute of Technology, funded by the National Aeronautics and Space Administration and the National Science Foundation. This research is based on observations with AKARI, a JAXA project with the participation of ESA.\\


\begin{thebibliography}{}

\bibitem[Alksnis et al. (2001)]{alksnis}Alksnis A., Balklavs A., Dzervitis U., Eglitis I., Paupers O., \& Pundure I., 2001, Baltic Astron., 10, 1

\bibitem[Brewer, Richer, \& Crabtree (1995)]{metal} Brewer J., Richer H., \& Crabtree D.\ 1995. \aj, 109, 2480

\bibitem[Bujarrabal, Fuente, \& Omont (1994)]{buj}Bujarrabal V., Fuente A., \& Omont A.\ 1994. \aap, 285, 247

\bibitem[Caswell et al. (1981)]{caswell} Caswell J. L., Haynes R. F., Goss W. M., \& Mebold U.\ 1981. Australian Journal of Physics, 34, 333

\bibitem[Chengalur et al.(1993)]{chengalur}Chengalur J. N., Lewis B. M., Eder J., \& Terzian Y., 1993, \apjs, 89, 189

\bibitem[Cho et al.(2016)]{ysosio}Cho S.-H., Yun Y., Kim J., Liu T., Kim K.-T., \& Choi M.\ 2016, \apj, 826, 157

\bibitem[Claussen, Kleinmann, Joyce, \& Jura (1987)]{claussen}Claussen, Kleinmann, Joyce, \& Jura 1987, \apjs, 65, 385

\bibitem[Cohen, Wheaton, \& Megeath (2003)]{2mass}Cohen, Martin; Wheaton, Wm. A.; Megeath, S. T. 2003, \aj, 126, 1090

\bibitem[Di Criscienzo et al. (2016)]{dicris}Di Criscienzo M.,  Ventura P.,  García-Hernández D. A.,  Dell'Agli F.,  Castellani M.,  Marrese P. M., Marinoni S.,  Giuffrida G.,  \& Zamora O. 2016, MNRAS 462, 395

\bibitem[Dinh-v-Trung \& Lim (2008)]{3lines}Dinh-V-Trung \& Lim J.\ 2008. \apj, 678, 303

\bibitem[Dunham et al. (2015)]{dunham}Dunham M. M., Allen L. E., Evans II N. J., et al.\ 2015,
\apjs, 220, 11

\bibitem[Egan et al. (2003)]{msxpsc}Egan, M. P., Price, S. D., Kraemer, K. E., et al. 2003, The
Midcourse Space Experiment Point Source Catalog, Version 2.3 (October 2003), Air Force Research Laboratory Technical Report AFRL-VS-TR-2003-1589 (Springfield, VA: NTIS)

\bibitem[Epchtein, Le Bertre, \& L\'epine(1990)]{epchtein} Epchtein, N., Le Bertre, T., \& Lepine, J.~R.~D.\ 1990, \aap, 227, 82

\bibitem[Fujii et al.\,(2006)]{fujii}Fujii T.,  Deguchi S.,  Ita Y.,  Izumiura H.,  Kameya O., Miyazaki A., \& Nakada Y. 2006, Publications of the Astronomical Society of Japan, 58, 529

\bibitem[Glass et al.\,(2001)]{glass} Glass I. S., Matsumoto S., Carter B. S., \& Sekiguchi K. 2001, \mnras, 321, 77

\bibitem[Guandalini et al. (2006)]{gua} Guandalini R., Busso M., Ciprini S., Silvestro G., \& Persi P.\ 2006, \aap, 445, 1069

\bibitem[Habing (1996)]{habing}Habing, H.~J\ 1988, \aap, 200, 40

\bibitem[Herwig (1996)]{herwig}Herwig, F.\ 2005, Annu. Rev. Atron. Astrophys., 43, 435

\bibitem[Hilburn (2018)]{eddie}Hilburn, E. "Infrared Properties of Stars in the Bulge Asymmetries and Dynamical Evolution Survey." 2018, https://digitalrepository.unm.edu/phyc\_etds/174

\bibitem[Ishihara et al.(2011)]{ishi} Ishihara D., Kaneda H., Onaka, T., Ita Y., Matsuura M., \& Matsunaga N.\ 2011, \aap, 534, A79

\bibitem[Kwok, Volk, \& Bidelman(1997)]{kvb} Kwok S., Volk K., \& Bidelman W.~P.\ 1997, \apjs, 112, 557

\bibitem[Le Bertre et al.\,(2003)]{lebertre}Le Bertre T., Tanaka M., Yamamura I., \& Murakami H.\,2003, \aap, 403, 943

\bibitem[Lindqvist et al.\,(1992)] {lindqv}Lindqvist M., Winnberg A., Habing H. J., \& Matthews H. E. 1992, \aap, 263, 183

\bibitem[Loup et al.\,(1993)]{loup} Loup C., Forveille T., Omont A., \& Paul J.~F.\ 1993, \aaps, 99, 291 

\bibitem[Lumsden et al.\,(2002)]{lumsden} Lumsden S.~L., Hoare M.~G., Oudmaijer R.~D., \& Richards D.\ 2002, \mnras, 336, 621

\bibitem[Matsunaga et al.\,(2009)]{matsunaga2} Matsunaga N., Kawadu T., Nishiyama S, Nagayama T., Hatano H, Tamura M., Glass I. S., \& Nagata T. 2009, \mnras 399, 1709

\bibitem[Matsunaga et al.\,(2017)]{matsunaga} Matsunaga N., Menzies J. W., Feast M. W., Whitelock P. A., Onozato H., Barway S., \& Aydi E.\,2017, \mnras, 469, 4949

\bibitem[Messineo et al.\,(2002)]{messineo} Messineo M.,  Habing H. J.,  Sjouwerman L. O.,  Omont A., \& Menten K. M. 2002, \aap, 393, 115

\bibitem[Nassau, Blanco, \& Morgan (1954)]{nassau}Nassau J. J., Blanco V. M., Morgan W. W. 1954, \apj, 120, 478

\bibitem[Noguchi, Aoki, \& Kawanomoto (2004)]{noguchi}Noguchi K, Aoki W, \& Kawanomoto S.\ 2004, \aap, 418, 67

\bibitem[Oloffson et al.\,(1981)]{olofsson} Olofsson H., Rydbeck O.E.H., Lane A. P., \& Predmore C. R.
1981, ApJL, 247, L81

\bibitem[Ortiz et al.\,(2005)]{ortiz} Ortiz, R., Lorenz-Martins, S., Maciel, W.~J., \& Rangel E.~M.\ 2005, \aap, 431, 565

\bibitem[Pihlstr\"om et al.\,(2018)]{ylva}Pihlstr\"om, Y. M., Sjouwerman, L. O., Claussen, M. J, Morris M. R., Rich R. M., van Langevelde H. J., \& Quiroga-Nuñez L.H. 2018, \apj, 868, 72


\bibitem[Rho et al.\,(2006)]{rho}Rho J., Reach W.T, Lefloch B., \& Fazio G. G. 2006, \apj, 643, 2, 965

\bibitem[Saito et al.\,(2012)]{saito}Saito R. K., Minniti D., Angeloni R., Catelan M. 2012, The Astronomer's Telegram, 4426

\bibitem[Samus et al.\,(2017)]{samus}Samus N. N., Kazarovets E. V., Durlevich O. V., Kireeva N. N., \& Pastukhova E. N. 2017, Astronomy reports, 61, 1, 80

\bibitem[Sevenster(2002)]{Sevenster2002} Sevenster, M.~N. \ 2002, \aj, 123, 2772

\bibitem[Sjouwerman, Capen, \& Claussen (2009)]{scc} Sjouwerman, L.~O., Capen, S.~M. \& Claussen, M.~J\ 2009, \apj, 705, 1554 (SCC09)

\bibitem[Sjouwerman, Messineo, \& Habing (2004)]{sjouwer}Sjouwerman L. O., Messineo M., \& Habing H. J.2004, Publications of the Astronomical Society of Japan, 56, 45

\bibitem[Sjouwerman et al.\,(2020)] {baade} Sjouwerman L. O. et al.\ 2020, in preparation

\bibitem[Soria-Ruiz et al.(2004)]{stype}Soria-Ruiz R., Alcolea J., Colomer F., Bujarrabal V., Desmurs J.-F., Marvel K. B.,\& Diamond P. J.\ 2004, \aap, 426, 131

\bibitem[Soszy\'nski et al.\,(2013)]{sos}Soszy\'nski I., Udalski A., Pietrukowicz P., Szyma\'nski M. K., Kubiak M., Pietrzy\'nski G., Wyrzykowski \L{}., Ulaczyk K., Poleski R., \& Koz\l{}owski S.\,2013, Acta Astronomica, 63, 1, 21

\bibitem[Stroh et al.(2018)]{michael}Stroh M.~C., Pihlstr\"om Y.~M., Sjouwerman L.~O., Claussen M.~J, Morris M.~R., \& Rich M.~R.\ 2018, \apj, 862, 14

\bibitem[Stroh et al.\,(2019)]{michael2}Stroh M.~C., Pihlstr\"om Y.~M., Sjouwerman L.~O., Lewis M.~O., Claussen M.~J, Morris M.~R., \& Rich M.~R.\  \,2019. \apj, 244, 2

\bibitem[Su et al.\,(2018)]{su}Su J. B., Shen Z.-Q., Chen X., \& Jiang D. R.\ 2018, \apj, 853, 42

\bibitem[Suh \& Kwon (2011)]{suh}Suh K. \& Kwon Y.\ 2011, \mnras, 417, 3047

\bibitem[Tanab\'e et al.\,(2008)]{akari}Tanabé, T. et al.\,2008, Publications of the Astronomical Society of Japan, 60, 375

 
\bibitem[te Lintel Hekkert et al.\,(1991)]{tlh}te Lintel Hekkert P., Caswell J. L., Habing H. J., et al. 1991, A\&AS, 90, 327

\bibitem[Tenenbaum et al.\,(2010)]{twostars}Tenenbaum E. D., Dodd J. L., Milam S. N., Woolf N. J., \& Ziurys L. M.\, 2010. \apj, 720, L102

\bibitem[Trapp et al.\,(2018)]{trapp}Trapp A., Rich R. M., Morris M. R., Sjouwerman L. O., Pihlström Y. M., Claussen M., \& Stroh M. C.\ 2018, \apj, 861, 14

\bibitem[Tsuji (1973)]{tsuji}Tsuji, T. 1973, \aap 23, 411

\bibitem[van der Veen \& Habing (1988)]{vdvhabing} van der Veen, W.~E.~C.~J., \& Habing, H.~J.\ 1988, \aap, 194,125 

\bibitem[van der Veen \& Rugers (1989)]{vdVR} van der Veen, W.E.C.J., \& Rugers, M.  1989. AA, 226, 183

\bibitem[Velilla Prieto et al.\,(2017)]{thatthingylvagaveme} Velilla Prieto L., S\'anchez Contreras C., Cernicharo J., Ag\'undez M., Quintana-Lacaci G., Bujarrabal V., Alcolea J., Balan\c{c}a C., Herpin F., Menten K. M., \& Wyrowski F.\, 2017. AA, 597, 46 

\bibitem[Xu et al. (2018)]{xu}Xu S.,  Zhang B.,  Reid M. J.,  Menten K. M.,  Zheng X.,  \& Wang G. 2018 ApJ, 859, 14

\end{thebibliography}
\end{document}